\documentclass[]{jfm}

\usepackage{graphicx}
\graphicspath{{figures/}} 
\usepackage{newtxtext}
\usepackage{newtxmath}
\usepackage{natbib}
\usepackage{multirow}
\usepackage{hyperref}
\hypersetup{
    colorlinks = true,
    urlcolor   = blue,
    citecolor  = black,
}

\newcommand{\RomanNumeralCaps}[1]

\newcommand{\bu}{{\bf u}}
\newcommand{\bb}{{\bf b}}
\newcommand\Pm{\mbox{\textit{Pm}}}
\newcommand\Rm{\mbox{\textit{Rm}}}  %
\renewcommand\Re{\mbox{\rm Re}}  %
\renewcommand\Im{\mbox{\rm Im}}  %

\newcommand{\kmax}{k_z^\mathrm{max}}

\title{Resistive instabilities in sinusoidal shear flows with a streamwise magnetic field}

\author{A.E.~Fraser\aff{1}, I.G.~Cresswell\aff{2}, \& P. Garaud\aff{1}}
\affiliation{\aff{1}Department of Applied Mathematics, Baskin School of Engineering, UC Santa Cruz, Santa Cruz CA 95064 
\aff{2}Department of Astrophysical and Planetary Sciences \& LASP, University of Colorado, Boulder, CO 80309, USA}

\begin{document}

\maketitle

\begin{abstract}
    We investigate the linear stability of a sinusoidal shear flow with an initially uniform streamwise magnetic field in the framework of incompressible magnetohydrodynamics (MHD) with finite resistivity and viscosity. This flow is known to be unstable to the Kelvin-Helmholtz instability in the hydrodynamic case. The same is true in ideal MHD, where dissipation is neglected, provided the magnetic field strength does not exceed a critical threshold beyond which magnetic tension stabilizes the flow. Here, we demonstrate that including viscosity and resistivity introduces two new modes of instability. One of these modes, which we call a resistively-unstable Alfv\'{e}n wave due to its connection to shear Alfv\'{e}n waves, exists for any nonzero magnetic field strength as long as the magnetic Prandtl number $\Pm < 1$. We present a reduced model for this instability that reveals its excitation mechanism to be the negative eddy viscosity of periodic shear flows described by \citet{DubrulleFrisch1991}. Finally, we demonstrate numerically that this mode saturates in a quasi-stationary state dominated by counter-propagating solitons. 
\end{abstract}


\section{Introduction}
\label{sec:intro}


The prevalence of shear in fluids makes it one of the most common sources of turbulence in nature. As such, interest in the linear and nonlinear stability of shear flows dates as far back as the 19th century to the early works of \citet{reynolds_1883} and \citet{helmholtz}, and still continues today. A large class of shear-driven instabilities can loosely be categorized as Kelvin-Helmholtz (KH) instabilities, which occur in  plane-parallel continuous or interfacial shear flows \citep{chandrasekhar,drazin}. Turbulence driven by KH instabilities can cause substantial mixing of momentum, heat and/or chemicals in fluid bodies such as the Earth's atmosphere, oceans, and liquid core, as well as planetary and stellar atmospheres and interiors. Quantifying shear-induced mixing is therefore a crucial step towards improving evolutionary models of these large-scale systems.


In fluids which are composed of partially or fully ionized plasma (e.g.~stellar interiors, magnetically-confined laboratory plasmas, etc.), or made of liquid metals (e.g.~planetary interiors), magnetic fields and the forces they exert must also be taken into account. Studies of the stability and mixing properties of parallel shear flows in that context are often performed using the magnetohydrodynamic (MHD) approximation \citep[see, e.g.][]{chandrasekhar,HughesTobias}, with a few notable exceptions \citep{rogers,karimabadi,henri,faganello,fraser2018,vogman}.
 
Many of these MHD works additionally use the ideal limit, where both viscosity and resistivity are neglected \citep[despite the fact that such mixing problems are often fundamentally ill-posed, see, e.g.,][]{Lecoanet2016}. In ideal MHD, the magnetic field lines are frozen into the flow and are forced to move with it. Meanwhile, a magnetic tension proportional to the square of the field amplitude resists the deformation of field lines, and has a tendency to rigidify the flow, imbuing it with elastic-like properties. 
As a result, the presence of a magnetic field parallel to the mean flow can hinder the development of KH billows. It has been shown that, in the ideal limit, a uniform streamwise magnetic field stabilizes KH modes provided its Alfv\'{e}n velocity exceeds the characteristic flow speed by a factor that depends on the flow profile but is typically of order unity \citep{chandrasekhar}.


Magnetic fields are also known to modify or fully invalidate several important exact theoretical results on the stability of incompressible, hydrodynamic,  parallel shear flows. For example, \citet{TatsunoDorland} showed that magnetized shear instabilities can exist even when the background flow does not have an inflection point, contrary to the hydrodynamic case where the latter is necessary \citep{Rayleigh}. Similarly, \citet{Lecoanet2010} provided examples of unstable magnetized stratified shear flows in which the Richardson number always exceeds $1/4$, showing that the Miles-Howard theorem \citep{miles,howard} does not apply in MHD. Finally, \citet{HughesTobias} showed that Howards' semicircle theorem \citep{howard} is modified in the presence of magnetic fields, and that the eigenvalues of the linear stability problem must now lie within the intersection of two semicircles in the complex plane, whose existence and position depend on the amplitude and profiles of the background flow and magnetic field.  


In non-ideal MHD, the resistivity of the fluid allows the field to partially decouple from the flow, and reduces its (usually) stabilizing influence. For magnetized KH instabilities, gradually increasing the resistivity can thus raise the growth rate of unstable modes to a value between that obtained in the ideal MHD limit and in the hydrodynamic case \citep{Palotti}. It is also worth mentioning that 
by contrast with the hydrodynamic case, three-dimensional perturbations are sometimes the fastest-growing modes in non-ideal MHD shear instabilities \citep{Hunt,HughesTobias}.

With these general results in mind, we investigate in this paper a very specific problem, namely the stability and evolution of a sinusoidal, incompressible shear flow with finite viscosity and resistivity, in the presence of a uniform, streamwise magnetic field. This problem is highly relevant to a number of applications but has not, to our knowledge, been studied in detail yet. Sinusoidal shear flows are defined here as unidirectional plane-parallel flows whose amplitude varies sinusoidally in the transverse direction. They commonly arise in nature from the development of a primary instability that results in the exponential growth of so-called "elevator" modes, as in homogeneous Rayleigh-B\'enard convection \citep[HRBC;][]{calzavarini,garaud_HRBC}, the double-diffusive fingering instability \citep{baines_thermohaline_1969}, and the Goldreich-Shubert-Fricke (GSF) instability \citep{Goldreich_Schubert,Fricke}. 
In these examples, secondary shear instabilities between elevators flowing in opposite directions are thought to be responsible for the saturation of the primary instability and have successfully been used to model it in the hydrodynamic limit \citep{RadkoSmith,Brown,Barker1}. In the MHD case, the presence of a uniform magnetic field aligned with the direction of the primary elevator mode flow has no effect on its growth rate, but can stabilize the secondary shearing mode, for the reasons discussed earlier. As such, understanding the linear stability and nonlinear evolution of magnetized shear instabilities in sinusoidal shear flows is a key step in quantifying the effects of magnetic fields on HRBC, and on various double-diffusive instabilities. Furthermore, numerical simulations of the latter 
show that the effective kinetic and/or magnetic Reynolds numbers of the saturated nonlinear flow can remain modest over a broad range of parameter space \citep{Brown}. As such, diffusive effects (viscosity and resistivity) should be taken into account to correctly model the development of the secondary shear instabilities.


In this paper, we therefore investigate the linear stability and nonlinear saturation of a sinusoidal, incompressible shear flow in the presence of a uniform, streamwise magnetic field. We present the background state and linearized equations in Section \ref{sec:model}. Section \ref{sec:linearresults} performs a linear stability analysis of this flow over a wide range of parameter space, demonstrating the presence of three distinct branches of instability, two of which have not, to our knowledge, been discussed before. We then focus on one of the two new branches, which exists even in the presence of very strong magnetic fields, and takes the form of overstable Alfv\'{e}n waves. We derive a heavily truncated model for the unstable modes in Section \ref{sec:reduced_model}, and use it in Section \ref{sec:physical_interpretation} to speculate that this instability is driven by the anti-diffusive properties of sinusoidal shear flows discussed by \citet{DubrulleFrisch1991}. In Section \ref{sec:simulations} we present an illustrative example of the  nonlinear evolution of this instability using a direct numerical simulation, demonstrating a linear growth phase consistent with our earlier linear stability analysis, and a saturated state dominated by counter-propagating solitons. We conclude in Section \ref{sec:conclusions} with a short discussion of the potential relevance of this new instability for natural systems, and of future work.

\section{Model and linear stability analysis}
\label{sec:model}

We consider a background flow ${\bf u}_E$ directed in the $z$ (streamwise) direction, whose amplitude varies sinusoidally in the $x$ (cross-stream) direction. Units are selected based on the flow's amplitude $U^*$ and horizontal wavenumber $k_x^*$, and in these units, ${\bf u}_E$ is given by 
\begin{equation}
\label{eq:u_E}
    {\bf u}_E = \sin(x) {\bf e}_z 
\end{equation}
(the subscript "$E$"  is used here to denote "elevator", following the motivating example given in Section \ref{sec:intro}). We assume that the background flow is maintained against viscous decay by an external force applied to the system, and is therefore a laminar steady-state solution of the governing equations (see below). We also assume the existence of a uniform background magnetic field $\mathbf{b}_E$ oriented in the streamwise direction, whose amplitude $B^*$ defines the unit magnetic field strength so that, in these units, 
\begin{equation}
\label{eq:b_E}
    \mathbf{b}_E = \mathbf{e}_z.
\end{equation}
The total flow and field are written as the sum of this background plus a perturbation, namely  
\begin{equation}
\label{eq:equilibrium_bg_decomp}
    {\bf u} = {\bf u}_E + \tilde{\bf u}, \quad {\bf b} = \mathbf{b}_E + \tilde{\bf b}, 
\end{equation}
and satisfy the governing equations
\begin{eqnarray}
\label{eq:normalized_full_PDEs}
\frac{\partial \bu}{\partial t} + \bu \cdot \bnabla \bu = - \nabla p  + C_B (\bnabla \times \bb) \times \bb +  \frac{1}{Re} \bnabla^2 (\bu-\bu_E),     \nonumber \\
\frac{\partial {\bf b}}{\partial t}  = \bnabla \times ({\bf u} \times {\bf b}) + \frac{1}{Rm} \bnabla^2 {\bf b}, \nonumber \\
\bnabla \cdot \bu = 0  \quad  \bnabla \cdot \bb = 0.
\end{eqnarray}
Note the viscous term in the momentum equation, where the non-dimensional applied force has been written as $-Re^{-1} \bnabla^2 {\bf u}_E$. The nondimensional parameters are the usual viscous and magnetic Reynolds numbers, as well as the ratio of characteristic magnetic and kinetic energies of the background flow (alternatively, an inverse Alfv\'enic Mach number squared), namely
\begin{equation}
    Re = \frac{U^*}{\nu^* k_x^*}, \quad  Rm = \frac{U^*}{\eta^* k_x^*} \mbox{  and   } C_B = \frac{(B^*)^2}{\rho_0^* \mu_0^* (U^*)^2},  
\end{equation}
where $\nu^*$ and $\eta^*$ are the kinematic viscosity and magnetic diffusivity of the fluid, $\rho_0^*$ is the constant density of the fluid and $\mu_0^*$ is the permeability of the vacuum. Note that the magnetic Prandtl number, $Pm = \nu^*/\eta^*$, is the ratio of $Rm$ and $Re$. It is usually smaller than one in stellar interiors, but not necessarily asymptotically small \citep{rincon_2019}. 

Linearization around the background state yields the evolution equations for the perturbations $\tilde{\bu}$ and $\tilde{\bb}$ as:
\begin{eqnarray}
\frac{\partial \tilde{\bu}}{\partial t} + \bu_E \cdot \bnabla \tilde{\bu} + \tilde{\bu} \cdot \bnabla \bu_E = - \nabla p  + C_B (\bnabla \times \tilde{\bb}) \times {\bf e}_z +  \frac{1}{Re} \bnabla^2 \tilde{\bu} ,     \nonumber \\
\frac{\partial \tilde {\bf b}}{\partial t}  = \bnabla \times ({\bf u}_E \times \tilde{\bf b})+ \bnabla \times (\tilde{\bf u} \times {\bf e}_z)  + \frac{1}{Rm} \bnabla^2 \tilde{\bf b}, \nonumber \\
\bnabla \cdot \tilde{\bu} = 0  \quad  \bnabla \cdot \tilde{\bb} = 0 , 
\label{eq:realspace-linsys}
\end{eqnarray}
which can be expressed, component-wise, as 
\begin{eqnarray}
\frac{\partial \tilde{u}_x}{\partial t} + \sin(x) \frac{\partial \tilde{u}_x}{\partial z} = - \frac{\partial p}{\partial x}  + C_B \left( \frac{\partial \tilde{b}_x}{\partial z} -\frac{\partial \tilde{b}_z}{\partial x}   \right) +  \frac{1}{Re} \nabla^2 \tilde{u}_x,    \nonumber \\
\frac{\partial \tilde{u}_y}{\partial t} + \sin(x) \frac{\partial \tilde{u}_y}{\partial z} =  - \frac{\partial p}{\partial y}  - C_B \left( \frac{\partial \tilde{b}_z}{\partial y} -\frac{\partial \tilde{b}_y}{\partial z}   \right)  +  \frac{1}{Re} \nabla^2 \tilde{u}_y ,     \nonumber \\
\frac{\partial \tilde{u}_z}{\partial t} + \sin(x) \frac{\partial \tilde{u}_z}{\partial z} + \tilde{u}_x \cos(x) = - \frac{\partial p}{\partial z} +  \frac{1}{Re} \nabla^2 \tilde{u}_z,   \nonumber \\
\frac{\partial \tilde{b}_x}{\partial t}  = - \sin(x) \frac{\partial \tilde{b}_x}{\partial z}+  \frac{\partial \tilde{u}_x}{\partial z}   + \frac{1}{Rm} \nabla^2 \tilde{b}_x, \nonumber \\
\frac{\partial \tilde {b}_y}{\partial t}  = - \sin(x) \frac{\partial \tilde{b}_y}{\partial z} +  \frac{\partial \tilde{u}_y}{\partial z}   + \frac{1}{Rm} \nabla^2 \tilde{b}_y, \nonumber \\
\frac{\partial \tilde {b}_z}{\partial t}  = - \sin(x) \frac{\partial \tilde{b}_z}{\partial z} + \cos(x) \tilde{b}_x +  \frac{\partial \tilde{u}_z}{\partial z}  + \frac{1}{Rm} \nabla^2 \tilde{b}_z, \nonumber \\
\frac{\partial \tilde{u}_x}{\partial x} + \frac{\partial \tilde{u}_y}{\partial y} + \frac{\partial \tilde{u}_z}{\partial z} = 0, \nonumber \\
\frac{\partial \tilde{b}_x}{\partial x} + \frac{\partial \tilde{b}_y}{\partial y} + \frac{\partial \tilde{b}_z}{\partial z} = 0, 
\end{eqnarray}
where we have defined $\tilde{\bf u}=(\tilde{u}_x,\tilde{u}_y,\tilde{u}_z)$ and $\tilde{\bf b}=(\tilde{b}_x,\tilde{b}_y,\tilde{b}_z)$. 

Next, we assume that the linearized eigenmodes have the same periodicity as that of the background flow (we have checked that these modes are the fastest-growing in all of the cases presented in this paper), and use the ansatz 
\begin{equation} \label{eq:FourierSeries}
\tilde{q}(x,y,z,t) =  \exp(ik_y y + i k_z z + \lambda t) \sum_{n=-\infty}^{\infty} q_n e^{inx},
\end{equation}
for $\tilde{q} \in \{ \tilde{u}_x,\tilde{u}_y,\tilde{u}_z,p,\tilde{b}_x,\tilde{b}_y,\tilde{b}_z\}$
to obtain the linear system:
\begin{eqnarray}
\lambda u_{x,n} +  \frac{k_z}{2}  (u_{x,n-1} - u_{x,n+1})   = - in p_n  + C_B \left( i k_z b_{x,n}- i n b_{z,n} \right) -  \frac{K_n^2}{Re} u_{x,n},   \nonumber \\
\lambda u_{y,n} +  \frac{k_z}{2}  (u_{y,n-1} - u_{y,n+1}) = -i k_y p_n  - C_B \left( i k_y b_{z,n}  -i k_z b_{y,n}  \right)  -  \frac{K_n^2}{Re} u_{y,n},      \nonumber \\
\lambda u_{z,n} +  \frac{k_z}{2}  (u_{z,n-1} - u_{z,n+1})+ \frac{1}{2}  (u_{x,n-1} + u_{x,n+1})  = - i k_z p_n  -  \frac{K_n^2}{Re} u_{z,n},   \nonumber \\
\lambda b_{x,n}  = -  \frac{k_z}{2} ( b_{x,n-1} - b_{x,n+1})  +  i k_z u_{x,n} -  \frac{K_n^2}{Rm} b_{x,n} ,\nonumber \\
\lambda b_{y,n}  = - \frac{k_z}{2} ( b_{y,n-1} - b_{y,n+1})   +  i k_z  u_{y,n}    -  \frac{K_n^2}{Rm} b_{y,n}, \nonumber \\
\lambda b_{z,n}  = - \frac{k_z}{2} ( b_{z,n-1} - b_{z,n+1})  + \frac{1}{2}  ( b_{x,n-1} + b_{x,n+1})   +  i k_z u_{z,n} -  \frac{K_n^2}{Rm} b_{z,n}, \nonumber \\
nu_{x,n} + k_y u_{y,n} + k_z u_{z,n} = 0 ,\nonumber \\
nb_{x,n} + k_y b_{y,n} + k_z b_{z,n} = 0 ,
\label{eq:linearsys}
\end{eqnarray}
where $K_n^2 = n^2 + k_y^2 + k_z^2$. Finally, by renaming $ip_n = \pi_n$, and $i b_{x,n} = \beta_{x,n}$ (and similarly for $b_{y,n}$ and $b_{z,n}$), the system can be cast into a form where all the coefficients are real. When truncated over a finite number of Fourier modes (so $n =-N,...,0,...,+N$), the first 7 equations form a generalized $7(2N+1) \times 7(2N+1)$ linear eigenvalue problem with constant real coefficients, which can be solved using standard linear algebra solvers (e.g. such as the LAPACK DGGEV routine). For each input wavenumber $(k_y,k_z)$, at a given set of input parameters $(Re,Rm,C_B)$, we select the eigenvalue which has the largest real part and refer to the latter as the growth rate of the mode $(k_y,k_z)$.

By contrast with the hydrodynamic case, the most unstable modes in MHD 
shear flows 
are not guaranteed to 
be strictly two-dimensional (with $k_y=0$) \citep{Hunt,vorobev}. However, for the parameter regimes explored here, we find that the growth rate of the 2D mode $(0,k_z)$ always exceeds that of corresponding 3D modes ($k_y,k_z$), for $k_y \ne 0$. In what follows, we therefore only discuss the properties of the 2D modes.

When the system is restricted to two-dimensional perturbations, the flow and field may more efficiently be expressed in terms of a streamfunction $\psi$ and flux function $A$, defined so that
\begin{eqnarray}
    \tilde{\bu} = \bnabla \times (\psi {\bf e}_y) = (-\partial \psi / \partial z, 0, \partial \psi / \partial x),  \\ 
    \tilde{\bb} = \bnabla \times (A {\bf e}_y)  = (-\partial A / \partial z, 0 , \partial A/\partial x).
\end{eqnarray}
With these definitions, the conditions $\bnabla \cdot \tilde \bu = 0$ and $\bnabla \cdot \tilde \bb =0$ are implicitly satisfied, and the linearized governing equations for $\psi$ and $A$ are: 
\begin{eqnarray}
\frac{\partial }{\partial t}(\nabla^2 \psi) + \sin(x) \frac{\partial}{\partial z} \nabla^2 \psi + \sin(x) \frac{\partial \psi}{\partial z} = C_B \partial_z \nabla^2 A + \frac{1}{Re} \nabla^4 \psi,  \label{eq:psieq}  \\ 
 \frac{\partial A}{\partial t}  =  \frac{\partial \psi}{\partial z}   - \sin(x) \frac{\partial A}{\partial z}  + \frac{1}{Rm} \nabla^2 A.  \label{eq:Aeq} 
\end{eqnarray}

Using ansatz~\eqref{eq:FourierSeries} as before, the linearized equations become
\begin{equation} \label{eq:psi_fourier_eq}
\lambda \psi_{n} = \frac{k_z}{2 K^2_{n}}\left[(1 - K^2_{n-1}) \psi_{n-1} - (1 - K^2_{n+1})\psi_{n+1}\right] + i C_B k_z A_{n} - \frac{1}{Re} K^2_{n} \psi_{n},
\end{equation}
and
\begin{equation} \label{eq:A_fourier_eq}
\lambda A_{n} = - \frac{1}{2} k_z \left( A_{n-1} - A_{n+1} \right) + i k_z \psi_{n} - \frac{1}{Rm}K^2_{n} A_{n}.
\end{equation}
This alternative definition of the flow and field, and its corresponding  equations, have been used to cross-check the results of the linear stability analysis for 2D modes, and will be useful in Sections \ref{sec:linearresults} and \ref{sec:reduced_model} below. Note that solving Eqs~\eqref{eq:psi_fourier_eq} and ~\eqref{eq:A_fourier_eq} is much faster than solving (\ref{eq:linearsys}).

\section{Results of the linear stability analysis and the existence of resistively-unstable Alfv\'{e}n waves for \texorpdfstring{$Pm < 1$}{Pm < 1}}
\label{sec:linearresults}
\begin{figure}
     \centering
     \includegraphics[width=\textwidth]{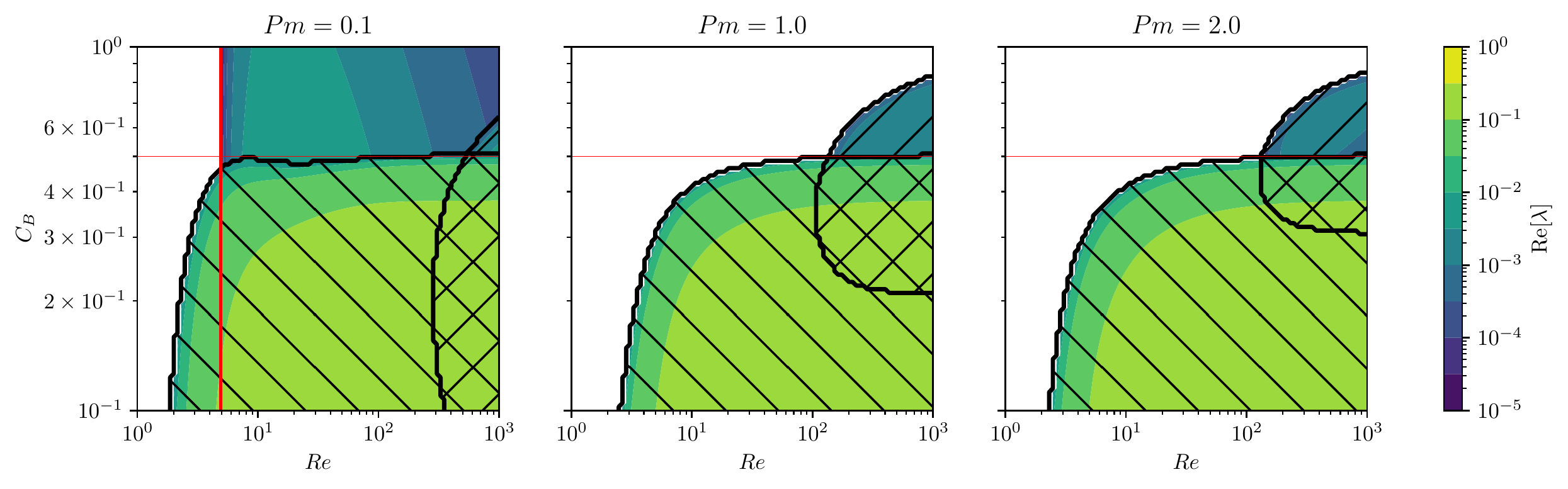}
     \caption{Growth rate of the fastest-growing  mode 
     as a function of $C_B$ and $\Rey$ for three values of $\Pm$. Black hatches indicate regions where the sinuous KH mode (hatches going down/to the right) and the varicose KH mode (hatches going down/to the left) are unstable. The red vertical line is the critical $\Rey$ for resistively-unstable Alfv\'{e}n waves calculated using Eq.~\eqref{eq:Re_crit}. 
     Red horizontal lines indicate $C_B = 0.5$, the marginal stability threshold in ideal MHD. 
     For $\Pm < 1$, resistively-unstable Alfv\'{e}n waves 
     exist for all $C_B$. 
     }
     \label{fig:growthrate_3Dscan}
 \end{figure}

For each set of physical parameters $C_B$, $\Rey$, and $\Rm$ (or, equivalently, $C_B$, $\Rey$, and $\Pm = \Rm/\Rey$), the stability of the equilibrium flow ${\bf u}_E = \sin(x) {\bf e}_z$ and field $\mathbf{b}_E = \mathbf{e}_z$ to 2D perturbations is assessed by solving Eqs.~\eqref{eq:psi_fourier_eq} and \eqref{eq:A_fourier_eq}  for the eigenvalues $\lambda$, for all possible wavenumbers $k_z$. If 
there exists a $k_z$ that admits one or more solutions with $\Re[\lambda] > 0$, then the system is said to be unstable 
at these parameters. 
When this is the case, we define $\kmax$ as the value of $k_z$ that maximizes $\Re[\lambda]$. Figure \ref{fig:growthrate_3Dscan} shows $\Re[\lambda(\kmax)]$ as a function of $C_B$ and $\Rey$ for three values of $\Pm$. For the parameters explored, we find that there are three distinct branches of instability in this system, which we describe in the following subsections.

 \begin{figure}
     \centering
     \includegraphics[width=\textwidth]{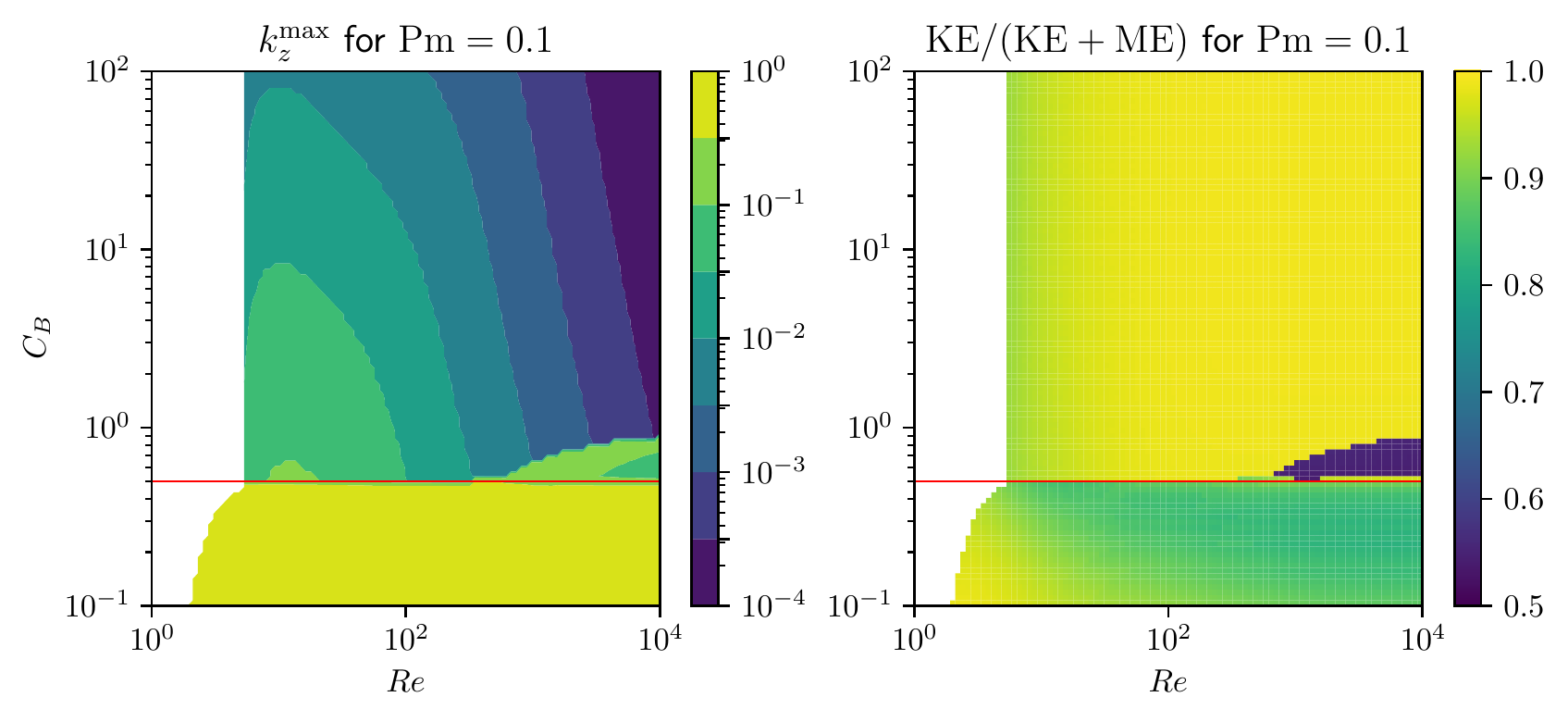}
     \caption{Left: Wavenumber of the fastest-growing mode, $\kmax$, as a function of $\Rey$ and $C_B$ for $\Pm = 1$. Right: Kinetic energy as a fraction of total energy for the fastest-growing mode. Note that the colorbar ranges from 0.5 to 1 -- the fraction never drops below 0.5 for the values shown here. Red horizontal lines indicate $C_B = 0.5$.}
     \label{fig:kmax_energy_scan}
 \end{figure}
 

\subsection{Sinuous KH modes}
\label{linearresults:subsec:sinuous}
The first of these three branches is a simple extension of the hydrodynamic KH instability, which continues to exist for sufficiently weak magnetic fields (small $C_B$). 
Regions in parameter space where this mode is unstable are marked by hatches going down and to the right in Fig.~\ref{fig:growthrate_3Dscan}.
We will refer to it as the ``sinuous" KH mode, because it meanders sideways, 
with a non-zero mean horizontal flow. Mathematically, this translates into a Fourier expansion [Eq.~\eqref{eq:FourierSeries}] where $\psi_0 \neq 0$ and (when $C_B \neq 0$) $A_0 \neq 0$.

The sinuous KH modes exist for sufficiently large $\Rey$ in the non-resistive case ($\Rm \to \infty$) for $C_B < 0.5$, i.e., as long as magnetic tension is small enough to permit the growth of KH billows. The modes also exist in the resistive case (finite $\Rm$), 
and in that case can persist at somewhat larger values of $C_B$ as long as $\Rm$ is 
low enough to relax the frozen-in-flux condition and reduce magnetic tension.

\begin{figure}
    \centering
    \includegraphics[width=\textwidth]{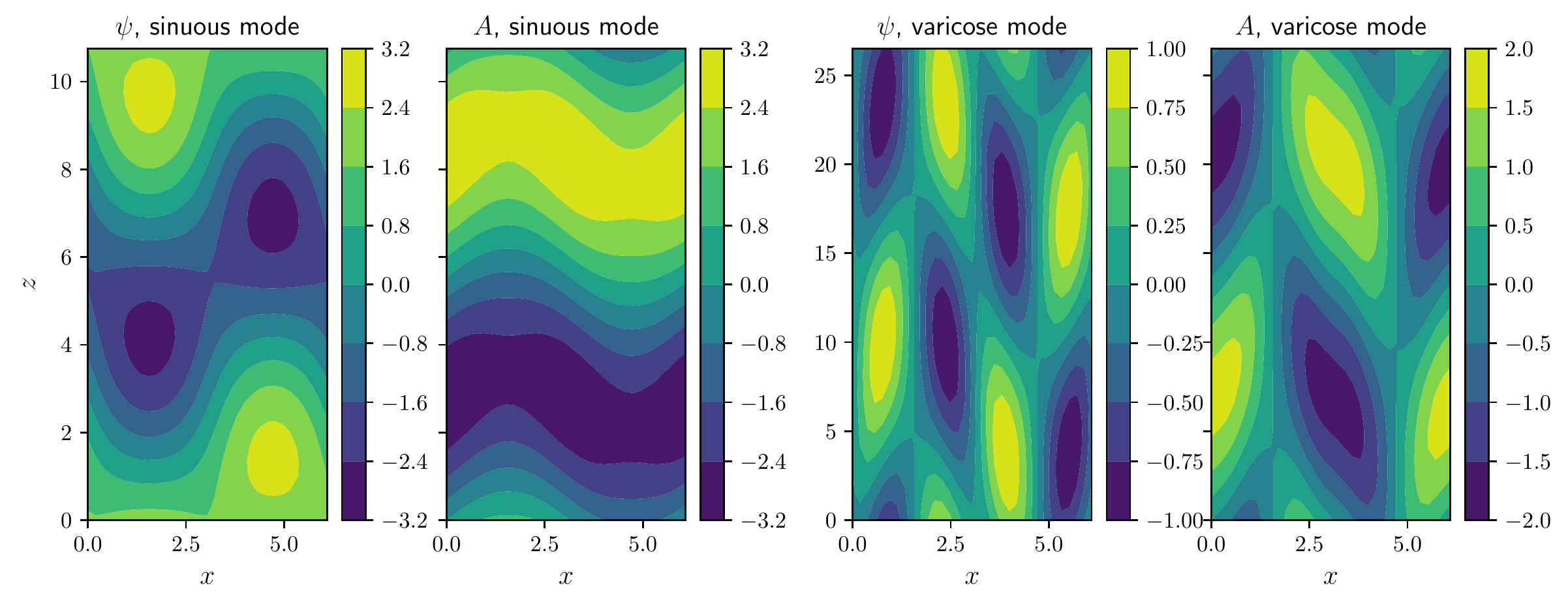}
    \caption{Mode structures are shown for the fastest-growing sinuous KH mode at $\Pm = 0.1$, $\Rey = 10^2$, $C_B = 0.1$ (left pair of panels), and the fastest-growing varicose KH mode at $\Pm = 0.1$, $\Rey = 10^3$, $C_B = 0.6$ (right pair). Within each pair of panels, the left panel shows contours of the streamfunction $\psi$, 
    and the right panel shows contours of the flux function $A$. Note that the (vertical) wavelength of each mode is $2 \pi / k_z^{\rm max}$, so the true aspect ratio of the modes is not accurately represented here.
    }
    \label{fig:KH_structures}
\end{figure}

The sinuous KH modes have $\Im[\lambda] = 0$, and generally have a growth rate $\Re[\lambda(\kmax)] \gtrsim~0.1$ for most of the physical parameters where they are found, except when they are nearly stabilized by either magnetic tension or dissipation. The wavenumber where their growth rate peaks is generally in the vicinity of $\kmax \sim 0.5$, as shown in Fig.~\ref{fig:kmax_energy_scan}. 
The two panels on the left in Fig.~\ref{fig:KH_structures} illustrate the structure of the fastest-growing mode ($k_z = \kmax$) for $C_B = 0.1$, $\Rey = 100$, and $\Pm = 0.1$. 
The first panel shows the contours of the streamfunction, 
representing streamlines of flow, and the second panel shows contours of the flux function, representing magnetic field lines. 
Here, $\psi$ and $A$ are the spatial structure of the eigenmodes, obtained by solving Eqs.~\eqref{eq:psi_fourier_eq} and \eqref{eq:A_fourier_eq} for $\psi_n$ and $A_n$, which are then used to compute $\psi = \Re[ \exp (i k_z z) \sum_n \psi_n e^{i n x} ]$ (and likewise for $A$). The amplitude is normalized such that the total energy of the mode (defined below) is $1$.
The structure of $\psi$, as expected, resembles that of a hydrodynamic KH mode. The perturbations take the form of alternating clockwise and counterclockwise recirculating eddies tilted against the mean shear. The dominant horizontal wavenumbers can be gleaned from the figure and include $n = 0$ (driving a mean horizontal flow, as discussed above) and $n = \pm 1$; 
higher-order wavenumbers are present as well, but not as prominent. 

The streamfunction and flux function of the eigenmodes can be used to compute their kinetic and magnetic energy, 
given in this nondimensionalization  by 
\begin{equation}
    \label{eq:KE_streamfunc_fourier}
    \mathrm{KE} = \frac{1}{2}\int dx \int dz \left[ \left( \frac{\partial \psi}{\partial z} \right)^2 + \left( \frac{\partial \psi}{\partial x} \right)^2 \right] = \sum_{n=-N}^N (k_z^2 + n^2) |\psi_n|^2,
\end{equation}
and
\begin{equation}
    \label{eq:ME_fluxfunc_fourier}
    \mathrm{ME} = \frac{C_B}{2}\int dx \int dz \left[ \left( \frac{\partial A}{\partial z} \right)^2 + \left( \frac{\partial A}{\partial x} \right)^2 \right] = C_B \sum_{n=-N}^N (k_z^2 + n^2) |A_n|^2.
\end{equation}
Using these definitions, we show in Fig.~\ref{fig:kmax_energy_scan} (right panel) the kinetic energy as a fraction of the total energy for the most unstable mode as a function of $\Rey$ and $C_B$ for $\Pm = 0.1$. Consistent with their interpretation as being primarily driven by a fundamentally hydrodynamic instability, we see that the kinetic energy of sinuous KH modes is significantly more than half of the total energy across these parameters.

\subsection{Varicose KH modes}
\label{linearresults:subsec:varicose}
Another class of unstable modes that also emerges is what we call the varicose KH modes hereafter, where the term varicose here is used by analogy with varicose modes in, e.g., shear instabilities in jets \citep{mattingly,drazin,mikhaylov}. While they also have $\Im[\lambda] = 0$, 
they can be distinguished from the sinuous KH modes because they have no  $x$-averaged horizontal flow or field, and thus $\psi_0 = A_0 = 0$. 

The varicose KH modes are only found at sufficiently large $\Rey$, but did not appear in ideal MHD in any of the cases we tested, nor do they exist in the hydrodynamic limit $C_B \rightarrow 0$. 
In Fig.~\ref{fig:growthrate_3Dscan}, the region of parameter space where varicose modes are unstable is marked by hatches going down and to the left. 
For $C_B < 0.5$, they are generally subdominant to sinuous modes; for $C_B \geq 0.5$, by contrast, varicose modes persist while sinuous modes are typically stabilized (for sufficiently large $\Rm$, as described in Sec.~\ref{linearresults:subsec:sinuous}). They are always eventually stabilized for sufficiently large magnetic field, however. 
Their most-unstable wavenumber, shown in Fig.~\ref{fig:kmax_energy_scan} (left), is generally on the order of $\kmax \sim 0.1$, slightly smaller than for the sinuous KH modes. Figure \ref{fig:kmax_energy_scan} (right) also shows that varicose modes are much closer to equipartition between kinetic and magnetic energy than sinuous modes are. 

Figure \ref{fig:KH_structures} shows that the structure of varicose modes differs significantly from sinuous modes. 
In particular, we see that the horizontal lengthscale of the perturbations is smaller, and dominated by the $n = \pm 2$ mode for the streamfunction, and the $n = \pm 1$ mode for the flux function. 
The flow contains convergent and divergent regions, consistent with varicose modes in other systems \citep{mattingly,drazin,mikhaylov}. 
While these modes appear (to our knowledge) to be unnoticed in the literature, they are not the focus of this paper, and will be discussed in greater detail in future work. 

\subsection{Resistively-unstable Alfv\'{e}n waves}
\label{linearresults:subsec:resistive}
 
\begin{figure}
    \centering
    \includegraphics[width=\textwidth]{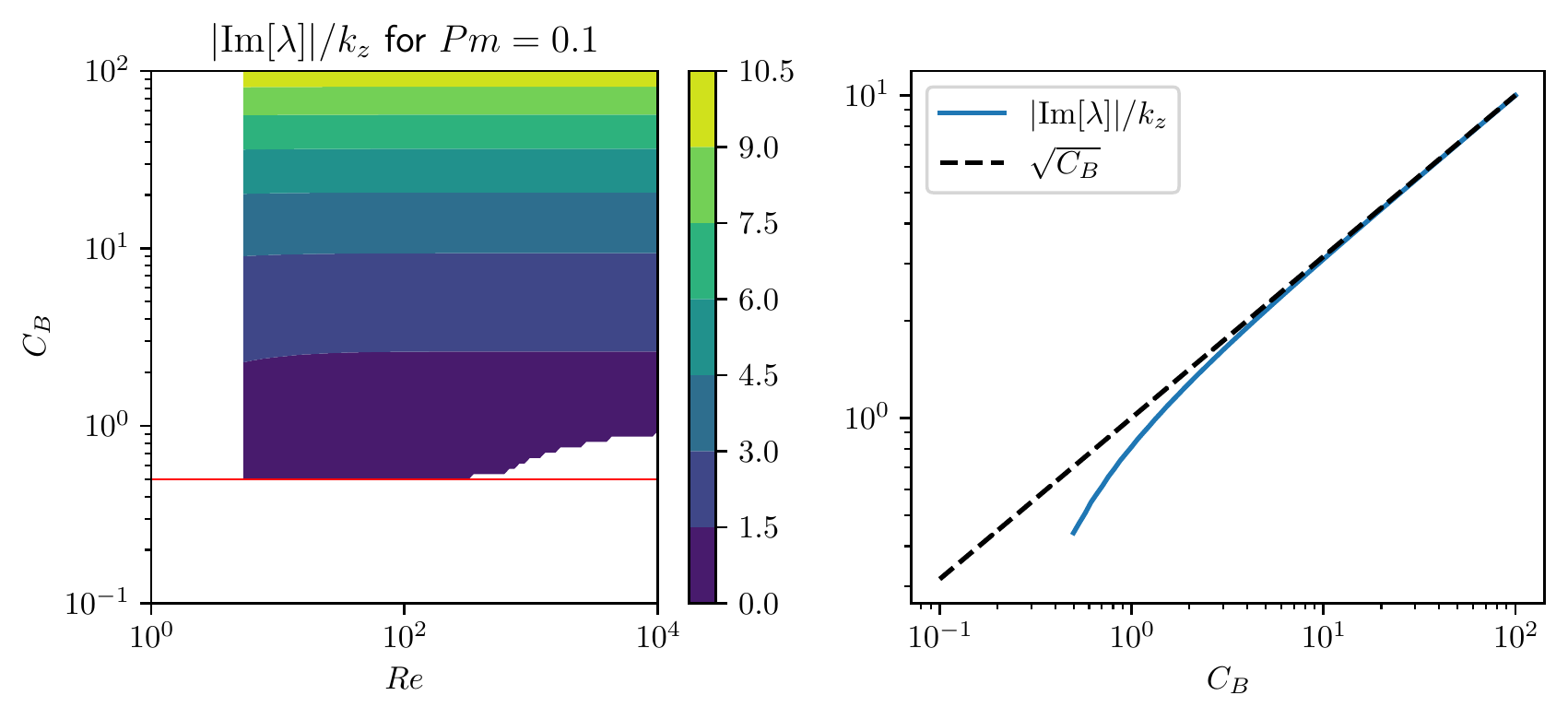}
    \caption{Left: Phase velocity $|\Im[\lambda]|/k_z$ of the fastest-growing mode at $\Pm = 0.1$ as a function of $C_B$ and $\Rey$, with the horizontal red line indicating $C_B = 0.5$. Right: Phase velocity versus $C_B$ for $\Rey = 100$, $Pm = 0.1$. Also shown is the nondimensional Alfv\'{e}n velocity $\sqrt{C_B}$ for reference.}
    \label{fig:phase_velocity}
\end{figure}

Contrary to the other modes discussed so far, the third type of unstable modes, called resistively-unstable Alfv\'{e}n waves hereafter, exist 
for arbitrarily large field strength $C_B$. They are found for $\Pm < 1$ and for $\Rey$ above a critical value that depends on $\Pm$. 
Unlike KH modes, they have nonzero frequencies $\Im[\lambda]$, and appear in complex-conjugate pairs at each unstable $k_z$. 
Their phase velocity $\Im[\lambda]/k_z$ scales roughly with the Alfv\'{e}n velocity $\sqrt{C_B}$, as shown in Fig.~\ref{fig:phase_velocity}. 
Figure \ref{fig:growthrate_3Dscan} shows that their growth rates are much smaller than those of KH modes, and decrease with increasing $C_B$. 
Meanwhile, Fig.~\ref{fig:kmax_energy_scan} reveals that the fastest-growing modes have much smaller values of $\kmax$ than KH modes, and that most of the energy is kinetic when $Pm\ll 1$. 

\begin{figure}
    \centering
    \includegraphics[width=\textwidth]{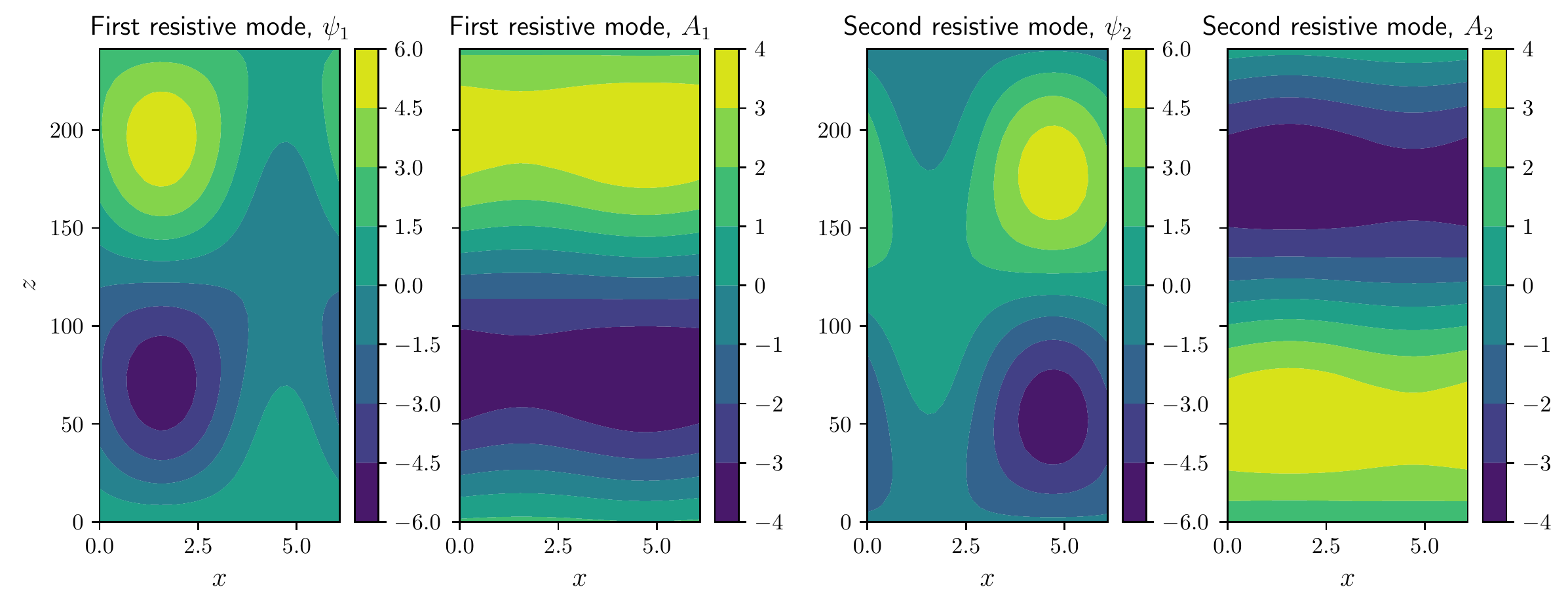}
    \caption{Mode structures are shown for the fastest-growing pair of resistively-unstable Alfv\'{e}n waves at $\Pm = 0.1$, $\Rey = 10^2$, $C_B = 1$, with each pair of panels showing contours of $\psi$ and $A$ as in Fig.~\ref{fig:KH_structures}. The left pair of modes is the one with $\Im[\lambda] > 0$, and thus travels in the $-z$ direction; the right pair of modes has $\Im[\lambda] < 0$ and travels in the $z$ direction. Note that the aspect ratio of the plots does not reflect the aspect ratio of the modes, whose vertical wavelength $2 \pi / k_z$ is much longer than $2 \pi$, the wavelength of the background flow.}
    \label{fig:resistive_structures}
\end{figure}

The structure of the fastest-growing resistively-unstable Alfv\'{e}n waves for $\Pm = 0.1$, $\Rey = 10^2$, and $C_B = 1$ is shown in Fig.~\ref{fig:resistive_structures}. 
We see that, contrary to both the sinuous and varicose KH modes, these 
unstable waves are highly asymmetric with respect to the background flow profile and behave differently depending on whether $\Im[\lambda] > 0$ (i.e.~the wave travels downward in the $-z$ direction) or $\Im[\lambda] < 0$ (i.e.~the wave travels upward). More specifically, we see that the downward-travelling perturbations are localized in the upward-moving region of the mean flow, and the upward-travelling perturbations are localized in the downward-moving region of the flow.

To our knowledge, the resistively-unstable Alfv\'{e}n waves have not been studied elsewhere in the literature. In what follows, we now present a reduced model of these new unstable modes in an effort to characterize them and clarify their origin.

\section{A reduced model for the resistively-unstable Alfv\'en waves}
\label{sec:reduced_model}

\begin{figure}
     \centering
     \includegraphics[width=\textwidth]{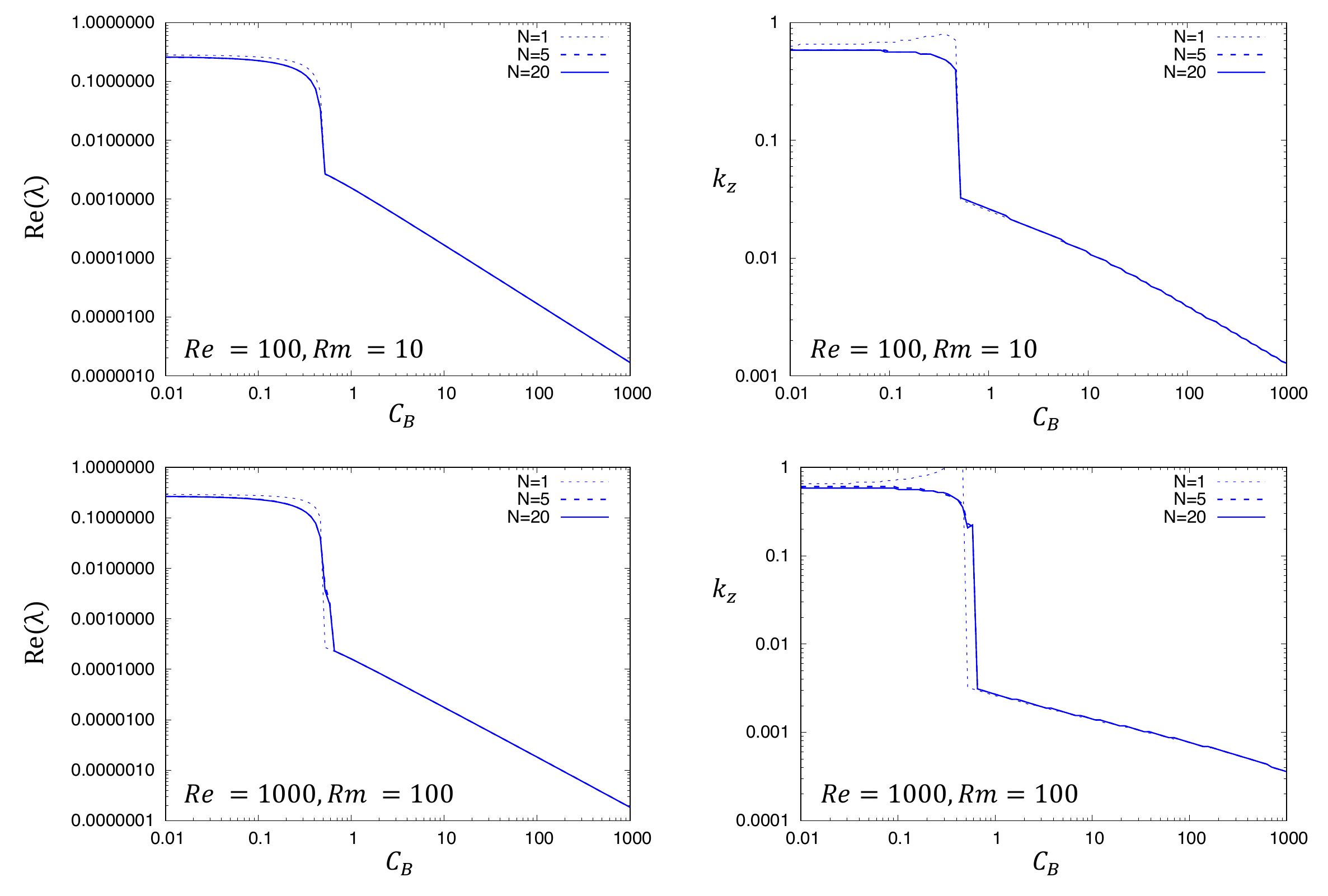}
     \caption{Growth rate $\Re(\lambda)$ (left) and wavenumber $k_{z}$ (right) of the fastest-growing mode of instability as a function of $C_B$, for $Re = 100$ (top) and $Re = 1000$ (bottom), for $N = 1$ (dotted line), $N=5$ (dashed line) and $N=20$ (solid line). In all cases $Pm = 0.1$. The $N=1$ truncation is an excellent approximation to the true system for the resistively-unstable Alfv\'{e}n waves.} 
     \label{fig:varyN}
 \end{figure}

We begin by noting that resistively-unstable Alfv\'en waves have a relatively simple spatial structure (especially at moderate Reynolds numbers) that is well-approximated by the most dramatic truncation of (\ref{eq:linearsys}), namely that for $N = 1$ (where the perturbations only contain a total of three Fourier modes, for $n = -1$, $n=0$ and $n=1$). This is illustrated in Figure \ref{fig:varyN}, which compares the growth rate and wavenumber of the fastest growing mode 
for truncations at $N = 20$, $N=5$ and $N=1$, respectively, for $Re = 100$ and $Re = 10000$. In all cases, $Pm = Rm/Re = 0.1$, and $C_B$ varies between 0.01 and 1000. We see that in general, the $N=1$ truncation captures most of the physics of the problem, including the overall amplitude of the growth rate $\Re(\lambda)$ and wavenumber $k_z$ of the fastest growing mode, and the clear regime transition that happens around $C_B = 0.5$. However, we also see that the properties of the resistively-unstable Alfv\'{e}n waves (which are the only modes that exist for $C_B \gg 0.5$) are particularly well captured by the $N=1$ truncation. 

With this in mind, we now consider the $N=1$ truncation only, and restrict our analysis to two-dimensional modes (so $k_y = \tilde{u}_y = 0$). We also limit our analysis to cases where $Pm< 1$ (so $Rm < Re$), as the modes of interest do not exist when $Pm \ge 1$. 

The $N=1$ truncation of the Fourier mode expansion is equivalent to seeking solutions of the kind
\begin{eqnarray}
    \psi(x,z,t) = e^{\lambda t + ik_z z} \left( \psi_0  +  \psi_1 e^{i x } + \psi_{-1} e^{- i x }  \right)  ,   \\ \label{eq:psidecomp}
    A(x,z,t) = i e^{\lambda t+ik_z z} \left( a_0   +  a_1 e^{i x } + a_{-1} e^{- i x }  \right) .   \label{eq:Adecomp}
\end{eqnarray}
Note that the added factor of $i$ in the expression for $A$ (so $ia_n = A_n$ for all $n$) makes the coefficients in the resultant system shown below real, which is both convenient and without loss of generality. The first term in each definition (terms in $\psi_0$, $a_0$) corresponds to $x-$invariant modes, sometimes referred to as "shearing modes" hereafter, while the second and third terms correspond to inclined modes that have structure in both horizontal and vertical directions. Substituting this ansatz into (\ref{eq:psieq}) and (\ref{eq:Aeq}), and projecting onto the relevant Fourier modes, yields the linear system
\begin{eqnarray}
&& \left( \lambda + \frac{k_z^2 }{Re} \right) \psi_0 -  \psi_E k_z   ( \psi_{1} - \psi_{-1})  =  -  k_z C_B a_0 ,  \nonumber \\ 
 &&  \left( \lambda +\frac{K^2 }{Re} \right) \psi_1 - \psi_E k_z \frac{1 - k_z^2}{ K^2}  \psi_0  = - k_z C_B   a_1  , \nonumber  \\ 
&&     \left( \lambda +  \frac{K^2}{Re} \right) \psi_{-1}  + \psi_E k_z \frac{1 - k_z^2}{ K^2} \psi_0 =  - k_z C_B  a_{-1} , \nonumber \\
&& \left( \lambda + \frac{k_z^2}{Rm} \right)  a_0 =  k_z \psi_0 -   \psi_E k_z   ( a_{-1} -  a_1 )  ,  \nonumber  \\ 
&&\left( \lambda + \frac{K^2}{Rm} \right)  a_1 =    k_z  \psi_1 -  \psi_E k_z  a_0 ,  \nonumber  \\ 
&& \left( \lambda + \frac{K^2 }{Rm} \right)  a_{-1} =   k_z  \psi_{-1}  + \psi_E k_z  a_0,
\label{eq:projected} 
\end{eqnarray}
 where here $K^2 = 1 + k_z^2$ and the scalar $\psi_E = 1/2$ is the amplitude of the Fourier coefficients of the streamfunction associated with the background shear flow ${\bf u}_E$. Its presence will later help identify the role of that shear flow in driving the resistive instability.  

Note that in the limit of extremely strong field and negligible diffusivity (i.e. neglecting terms in $\psi_E$ and in $1/Re$ or $1/Rm$), the system reduces to
\begin{eqnarray}
&& \lambda  \psi_0  =  - k_z C_B  a_0, \quad \lambda  a_0 = k_z \psi_0,  \nonumber  \\ 
 &&  \lambda \psi_1   = - k_z C_B   a_1, \quad  \lambda a_1 =   k_z  \psi_1,  \nonumber  \\ 
&&  \lambda \psi_{-1}  = - k_z C_B a_{-1}, \quad \lambda  a_{-1} =  k_z  \psi_{-1}, 
\end{eqnarray}
so that $\lambda^2 q = - k_z^2 C_B q$ for $q \in \{ \psi_0, \psi_1, \psi_{-1}, a_0, a_1, a_{-1} \}$. Each of these leads to $\lambda = \pm i k_z \sqrt{C_B}$, which is the non-dimensional version of the dispersion relationship for non-dissipative Alfv\'en waves travelling on a constant vertical magnetic field, which is as expected in the limit considered.   

Going back to (\ref{eq:projected}) and taking the sum of the $\psi_1$ and $\psi_{-1}$ equations, and the sum of the $a_{-1}$ and $a_1$ equations, we obtain the reduced system 
\begin{eqnarray}
     \left( \lambda +  \frac{K^2 }{Re} \right) (\psi_1 + \psi_{-1})   = - k_z C_B   ( a_1 + a_{-1}),  \nonumber  \\ 
     \left( \lambda +  \frac{K^2 }{Rm} \right) ( a_1 + a_{-1})  =  k_z   (\psi_1 + \psi_{-1}),   
\end{eqnarray}
which shows that, for modes with $a_1 + a_{-1}\neq0$ and $\psi_1 + \psi_{-1} \neq 0$, $\lambda$ satisfies the simple quadratic equation 
\begin{equation}
     \left( \lambda +  \frac{K^2 }{Re} \right)\left( \lambda +  \frac{K^2 }{Rm} \right) = - k_z^2 C_B,
     \label{eq:fastA}
    \end{equation}
whose solutions are 
\begin{equation}
\lambda = - \frac{K^2}{2} \left( \frac{1}{Re} + \frac{1}{Rm} \right) \pm  i k_z \sqrt{C_B} \sqrt{ 1- \frac{K^4}{4k_z^2 C_B} \left( \frac{1}{Re} - \frac{1}{Rm} \right)^2  }. 
\end{equation}
These correspond to viscously and resistively damped Alfv\'en waves, that decay on the timescale of order $Rm/K^2$, with $K$ of order unity. This result implies that the system dynamics relax to the subset of eigenmodes for which $a_1 =-a_{-1}$, and $\psi_1 = - \psi_{-1}$, on a relatively short timescale (unless $Rm$ is very large).

To look outside of this strictly decaying subspace (since we are looking for growing modes), we then assume $a_1 =-a_{-1}$, and $\psi_1 = - \psi_{-1}$ in  (\ref{eq:projected}), and obtain the new reduced system 
\begin{eqnarray}
&& \left( \lambda +  \frac{k_z^2 }{Re} \right) \psi_0 = 2 \psi_E k_z \psi_{1}   -  k_z C_B a_0 ,  \nonumber \\ 
&&   \left( \lambda + \frac{K^2}{Re} \right) \psi_{1}  =   \psi_E k_z \frac{1 - k_z^2}{ K^2} \psi_0   - k_z C_B  a_1  , \nonumber \\
&& \left( \lambda + \frac{k_z^2}{Rm}\right) a_0 =  k_z \psi_0 +   2 \psi_E k_z   a_1 ,  \nonumber  \\ 
&& \left( \lambda + \frac{K^2 }{Rm} \right) a_1 =    k_z  \psi_{1}  - \psi_E k_z  a_0.
\label{eq:projected_small} 
\end{eqnarray}
It can be shown with a little algebra that this yields a quartic equation for the eigenvalue $\lambda$:
\begin{eqnarray}
\left(\lambda + \frac{K^2}{Re}\right)\left(\lambda+ \frac{ k_z^2}{Re}\right)\left(\lambda+ \frac{K^2}{Rm}\right)\left(\lambda + \frac{ k_z^2}{Rm} \right)\nonumber  \\
+ k^2_z C_B \left[ \left(\lambda+ \frac{K^2}{Rm}\right) \left(\lambda + \frac{K^2}{Re}\right) + \left(\lambda+ \frac{ k_z^2}{Re}\right)\left(\lambda + \frac{ k_z^2}{Rm} \right)   \right] \nonumber\\ 
- 2\psi_E^2 k_z^2 \frac{1 - k_z^2}{ K^2}  \left(\lambda+ \frac{K^2}{Rm}\right) \left(\lambda+ \frac{k_z^2}{Rm}\right)  + 2\psi_E^2k_z^2  \left(\lambda + \frac{K^2}{Re}\right)\left(\lambda + \frac{k_z^2}{Re}\right)\nonumber\\
= 4\psi_E^2 k^4_z C_B  \frac{k_z^2}{ K^2} - (k^4_z C_B^2 - 4\psi_E^4 k_z^4 \frac{1 - k_z^2}{ K^2}  ).  \label{eq:disprel}
\end{eqnarray} 

Based on the finding that the resistively-unstable Alfv\'{e}n waves dominate only for large magnetic field strength (i.e. $C_B > 1/2$), but persist at arbitrarily large values of $C_B$, we postulate that they must somehow draw their energy from the interaction between other more weakly damped Alfv\'en modes and the background sinusoidal shear flow. For this reason we now rescale $\lambda$ with the Alfv\'en frequency $\sqrt{C_B} k_z$, introducing the variable $\hat \lambda = \lambda /(k_z \sqrt{C_B})$. We found in Figure \ref{fig:phase_velocity} that the imaginary part of $\hat \lambda = O(1)$ for the resistively-unstable Alfv\'{e}n waves. With this rescaling, we obtain
\begin{eqnarray}
\left(\hat \lambda + \frac{K^2}{Re\sqrt{C_B}k_z}\right)\left(\hat \lambda+ \frac{ k_z}{Re\sqrt{C_B}}\right)\left(\hat \lambda+ \frac{K^2}{Rm\sqrt{C_B}k_z}\right)\left(\hat \lambda + \frac{ k_z}{Rm\sqrt{C_B}} \right)\nonumber  \\
+  \left(\hat \lambda+ \frac{K^2}{Rm\sqrt{C_B}k_z}\right) \left(\hat \lambda + \frac{K^2}{Re\sqrt{C_B}k_z}\right) + \left(\hat \lambda+ \frac{ k_z}{Re\sqrt{C_B}}\right)\left(\hat \lambda + \frac{ k_z}{Rm\sqrt{C_B}} \right)   \nonumber\\ 
- 2\frac{\psi_E^2}{C_B}  \frac{1 - k_z^2}{ K^2}  \left(\hat \lambda+ \frac{K^2}{Rm\sqrt{C_B}k_z}\right) \left(\hat \lambda+ \frac{k_z}{Rm\sqrt{C_B}}\right) \nonumber \\ + 2\frac{\psi_E^2}{C_B} \left(\hat \lambda + \frac{K^2}{Re\sqrt{C_B}k_z}\right)\left(\hat \lambda + \frac{k_z}{Re\sqrt{C_B} }\right)\nonumber\\
= -1 + 4 \frac{\psi_E^2}{C_B}  \frac{k_z^2}{ K^2} + 4\frac{\psi_E^4}{C_B^2} \frac{1 - k_z^2}{ K^2} .   \label{eq:disprel}
\end{eqnarray} 
This demonstrates the emergence of two types of constant terms: those in $\psi_E^2 / C_B$, which are proportional to the square of the Alv\'enic Mach number associated with the background flow and field, and those in $K^2/Re \sqrt{C_B}k_z$, $K^2/Rm \sqrt{C_B}k_z$, $k_z/Re\sqrt{C_B}$ or $k_z/Rm\sqrt{C_B}$, which are the ratio of a viscous or magnetic diffusion rate to the Alfv\'en oscillation rate. 
At a fixed mean flow amplitude (as is necessarily implied in the non-dimensionalization selected here) and a fixed mode structure (i.e. fixed $k_z$), all of these terms go to zero as the magnetic field strength ($C_B$) increases. However, had we selected a different non-dimensionalization, it would be possible to let the term in $\psi_E^2$ go to zero independently of the diffusive terms. 
It is this second route that we now take here, as it leads to the correct model for the resistively-unstable Alfv\'{e}n waves. 

For this reason, we now introduce the small parameter $\epsilon = 2\frac{\psi_E^2}{C_B}$, which we take to be small, without necessarily requiring that the diffusive terms be small. This apparent inconsistency would be resolved in a different non-dimensionalization, but at the cost of introducing a new system of units and new notations, which we prefer to avoid. In addition, the diffusive terms all contain the quantity $k_z$, whose size relative to $\epsilon$ can vary. As such, we cannot a priori neglect them.  

We then assume an asymptotic expansion of the kind $\hat \lambda = \hat \lambda_0 + \epsilon \hat \lambda_1$ (which will be verified a posteriori to be correct). To study what happens in the {\it absence} of a mean flow, we artificially set $\psi_E = 0$, or equivalently, $\epsilon = 0$. In this limit, the equation reduces to two possible quadratic equations in $\hat \lambda_0$, namely the equations for the decay rates of diffusive Alfv\'en waves:  
\begin{eqnarray}
 \left(\hat \lambda_0+ \frac{K^2}{Rm\sqrt{C_B}k_z}\right) \left(\hat \lambda_0 + \frac{K^2}{Re\sqrt{C_B}k_z}\right) + 1 = 0,  \nonumber  \\ 
\left(\hat \lambda_0 + \frac{ k_z}{Re\sqrt{C_B}}\right)\left(\hat \lambda_0 + \frac{ k_z}{Rm\sqrt{C_B}} \right)  \label{eq:eqlambda0}
+ 1 = 0.
\end{eqnarray}
The first one is identical to (\ref{eq:fastA}), and was discarded on the basis that the corresponding modes decay too rapidly. We continue to discard it here, as it would lead to modes that do not grow. The second one has solutions that decay on the timescale $O(Re/k^2_z)$, which can be very long provided $k_z$ is sufficiently small: 
\begin{equation}
\hat \lambda_0 = - \frac{ k_z}{2\sqrt{C_B}  } \left(\frac{1}{Re} + \frac{1}{Rm}\right) \pm i \sqrt{ 1- \frac{ k_z^2}{4C_B}  \left(\frac{1}{Re}- \frac{1}{Rm}\right)^2 }. 
\label{eq:lambda0}
\end{equation}
These modes are pure "shearing" Alfv\'en modes in the terminology introduced earlier, i.e, at lowest order their corresponding velocity field is invariant in the $x$-direction. As we shall demonstrate, it is the interaction of these modes with the background shear that drives the growth of resistively-unstable Alfv\'{e}n waves.

\begin{figure}
     \centering
     \includegraphics[width=\textwidth]{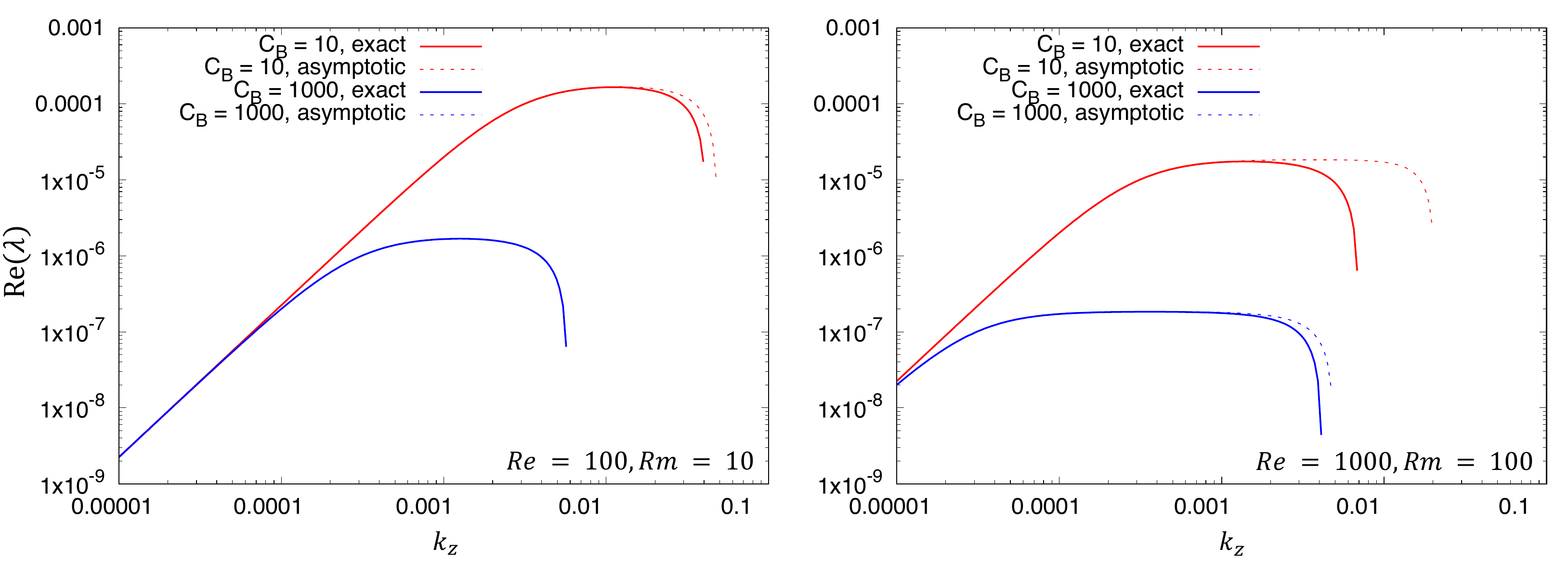}
     \caption{Growth rate $\Re(\lambda)$ as a function of $k_z$ in the reduced model ($N=1$ truncation), for $Re=100$ and $Rm=10$ (left) and $Re=1000$ and $Rm=100$ (right). In both cases, two values of $C_B$ are shown: $C_B = 10$ (red) and $C_B = 1000$ (blue). The exact solution (solid line) is obtained numerically, the asymptotic solution (dotted line) is obtained using equations (\ref{eq:lambda0}) and (\ref{eq:lambda1eq}).}
     \label{fig:asymp1}
 \end{figure}

Expanding (\ref{eq:disprel}) to first order in $\epsilon$ (but keeping the diffusive terms whose size is unknown without specifying $k_z$), and using (\ref{eq:eqlambda0}) and (\ref{eq:lambda0}), we find that the first-order correction      $\hat \lambda_1$ satisfies the linear equation
\begin{eqnarray}
 \hat \lambda_1 \left[ \left(\hat \lambda_0+\frac{ K^2}{Rm \sqrt{C_B} k_z }\right)\left( \hat \lambda_0 + \frac{K^2}{Re\sqrt{C_B}  k_z}\right) + 1 \right] \left[ 2 \hat \lambda_0 + \frac{k_z}{\sqrt{C_B}} \left(\frac{1}{Re} + \frac{1}{Rm}\right)   \right] \nonumber  \\ 
=  2\frac{k_z^2}{K^2}  +  \frac{1-k_z^2}{K^2}  \left( \hat \lambda_0+ \frac{ K^2}{Rm \sqrt{C_B} k_z }\right) \left( \hat \lambda_0 + \frac{ k_z}{Rm \sqrt{C_B}}\right)  \nonumber \\ -  \left( \hat \lambda_0+ \frac{k_z}{Re\sqrt{C_B} }\right)\left( \hat \lambda_0 + \frac{ K^2}{Re\sqrt{C_B} k_z }\right).
\label{eq:lambda1eq}
\end{eqnarray}  
Figure \ref{fig:asymp1} shows the real part of the asymptotic solution $\lambda = k_z \sqrt{C_B} ( \hat \lambda_0 + \epsilon \hat \lambda_1)$, with $\hat \lambda_0$ given by (\ref{eq:lambda0}) and $\hat \lambda_1$ given by the solution of (\ref{eq:lambda1eq}), and compares it with the growth rate obtained from the numerical solution of (\ref{eq:linearsys}) computed using the $N = 1$ truncation, for $k_y = 0$. Several values of the input parameters $Re$, $Rm$ and $C_B$ are tested. 

First and foremost, we confirm that there are indeed solutions with a positive real part for small enough $k_z$ in this particular subspace of the highly reduced model, demonstrating that the resistive instability does occur through the interaction of the weakly decaying "shearing" Alv\'en mode with the mean sinusoidal flow, as suspected. 
We see that the asymptotic expression is always appropriate for sufficiently large $C_B$, again as expected, but we also see that the expansion is not uniform in $Re$ and $k_z$. In particular, the approximation seems to always be valid for sufficiently small $k_z$ even when $C_B$ is not very large, but begins to fail at larger $k_z$, and does so earlier for larger Reynolds numbers $Re$ and $Rm$. The reason for the non-uniformity of the asymptotic expansion will be clarified later.

In what follows, we now aim to obtain a fully analytical solution to the problem, at least in some region of parameter space, that will allow us to gain a better understanding of the nature of the instability. To do so, we first limit the analysis to the range of parameters for which the asymptotic approximation to first order in $\epsilon$, namely equation (\ref{eq:lambda1eq}), is valid. We also capitalize on the fact that the resistively-unstable Alfv\'{e}n waves have a very small wavenumber (see Section \ref{sec:linearresults}), and further expand the solution in the limit of $k_z \rightarrow 0$. In what follows we assume (and later justify) that $k_z$ is $o(Re^{-1})$.  
 
When this is the case, and keeping only terms of lowest order in $k_z$, we have 
\begin{equation}
\hat \lambda_0 \simeq  - \frac{ k_z}{2\sqrt{C_B}  } \left(\frac{1}{Re} + \frac{1}{Rm}\right) \pm i, 
\label{eq:lambda0_approx}
\end{equation}
and after some cumbersome but otherwise straightforward algebra, again keeping only  low-order terms in $k_z$ and/or $Re^{-1}$, we obtain the expression
 \begin{equation}
    \hat \lambda_1 \simeq \frac{(Re-Rm) \sqrt{C_B} k_z}{2K^2 \left[ 1   + \frac{C_B  k^2_z }{K^4}  \left(Re + Rm \right)^2  \right] }  \left[ 1   \mp i \frac{\sqrt{C_B}  k_z }{K^2}  \left(Re + Rm \right)  \right] .
    \label{eq:lambda1_approx}
\end{equation} 


For the Taylor expansion of $\hat \lambda$ in the small parameter $\epsilon$ to be meaningful, one needs both $\epsilon |\Re(\hat \lambda_1)| \ll |\hat \lambda_0|$ and $\epsilon \Im(\hat \lambda_1) \ll |\hat \lambda_0|$. With $|\hat \lambda_0| \sim O(1)$, and noting that the second constraint is more stringent than the first, we obtain the condition
\begin{equation}
k^2_z  \ll \frac{2K^2}{(Re^2-Rm^2)} \rightarrow k_z = o(Re^{-1}),
\end{equation}
as discussed above (assuming $Rm$ is not too close to $Re$, and noting that $K^2 =  O(1)$). For values of $k_z$ approaching $O(Re^{-1})$ the expansion is no longer strictly valid, but remains adequate, which explains the trends seen in Figure \ref{fig:asymp1}. Finally, we can substitute this expression and the one for $\hat \lambda_0$ into $\lambda = k_z \sqrt{C_B} ( \hat \lambda_0 + \epsilon \hat \lambda_1)$ to obtain the real part of $\lambda$: 
\begin{eqnarray}
\Re(\lambda) \simeq  - \frac{  k_z^2}{2  } \left(\frac{1}{Re} + \frac{1}{Rm}\right) 
+  \frac{k_z^2 \psi_E^2}{K^2}\frac{Re-Rm}{ 1 +  \frac{C_B k^2_z}{K^4}  (Rm + Re )^2  }  \cdot  \label{eq:asymp_simple}
 \end{eqnarray}  

\begin{figure}
      \centering
      \includegraphics[width=0.7\textwidth]{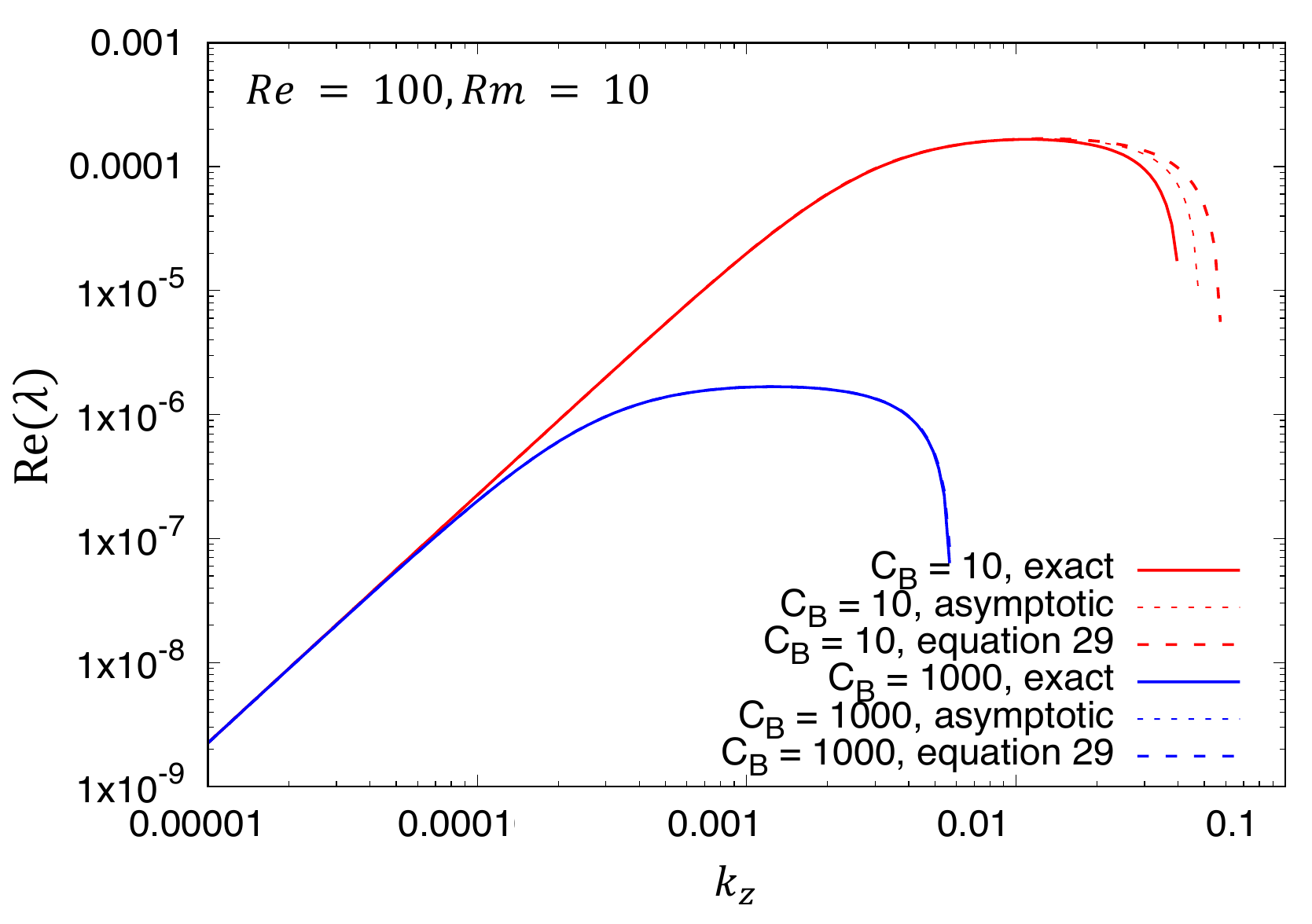}
      \caption{Growth rate as a function of $k_z$ in the reduced model, for $Re=100$ and $Rm=10$. The exact solution (solid line) is obtained numerically, the asymptotic solution (dotted line) is obtained using equations (\ref{eq:lambda0}) and (\ref{eq:lambda1eq}), and the dashed line is the analytical expression (\ref{eq:asymp_simple}).}
      \label{fig:asymp2}
  \end{figure}

This expression reveals that $\Re(\lambda)$ can only be positive when $Re > Rm$, or in other words, when $Pm < 1$, a result that is consistent with our findings from Section \ref{sec:linearresults}. 
We can also see in Figure \ref{fig:asymp2} that equation (\ref{eq:asymp_simple})
is a good asymptotic approximation to $\Re(\lambda)$ for all $k_z$ in the limit of large $C_B$ and moderate $Re$ and $Rm$ (e.g. the case $Re = 100$, $Rm = 10$, and $C_B = 1000$), and a reasonable approximation even for smaller $C_B$.  
 
%
 
 
 This analytical expression can be used to deduce some of the salient properties of the instability. For example, a criterion for instability can be obtained by requiring that $\Re(\lambda)>0$ as $k_z \rightarrow 0$, and noting that $\psi_E = 1/2$ and $K^2 \simeq 1$, we find that unstable modes exist provided: 
 \begin{equation}
 ReRm \frac{Re-Rm}{Re+Rm} > 2,
 \end{equation}
 or equivalently, 
  \begin{equation}
 Re  > \sqrt{\frac{2}{Pm} \frac{1+Pm}{1-Pm}} \quad \mbox{ (for $Pm < 1$)}.
 \label{eq:Re_crit}
 \end{equation} 
This shows that the instability is suppressed as $Pm \rightarrow 1$ from below. 
 
\begin{figure}
      \centering
      \includegraphics[width=0.7\textwidth]{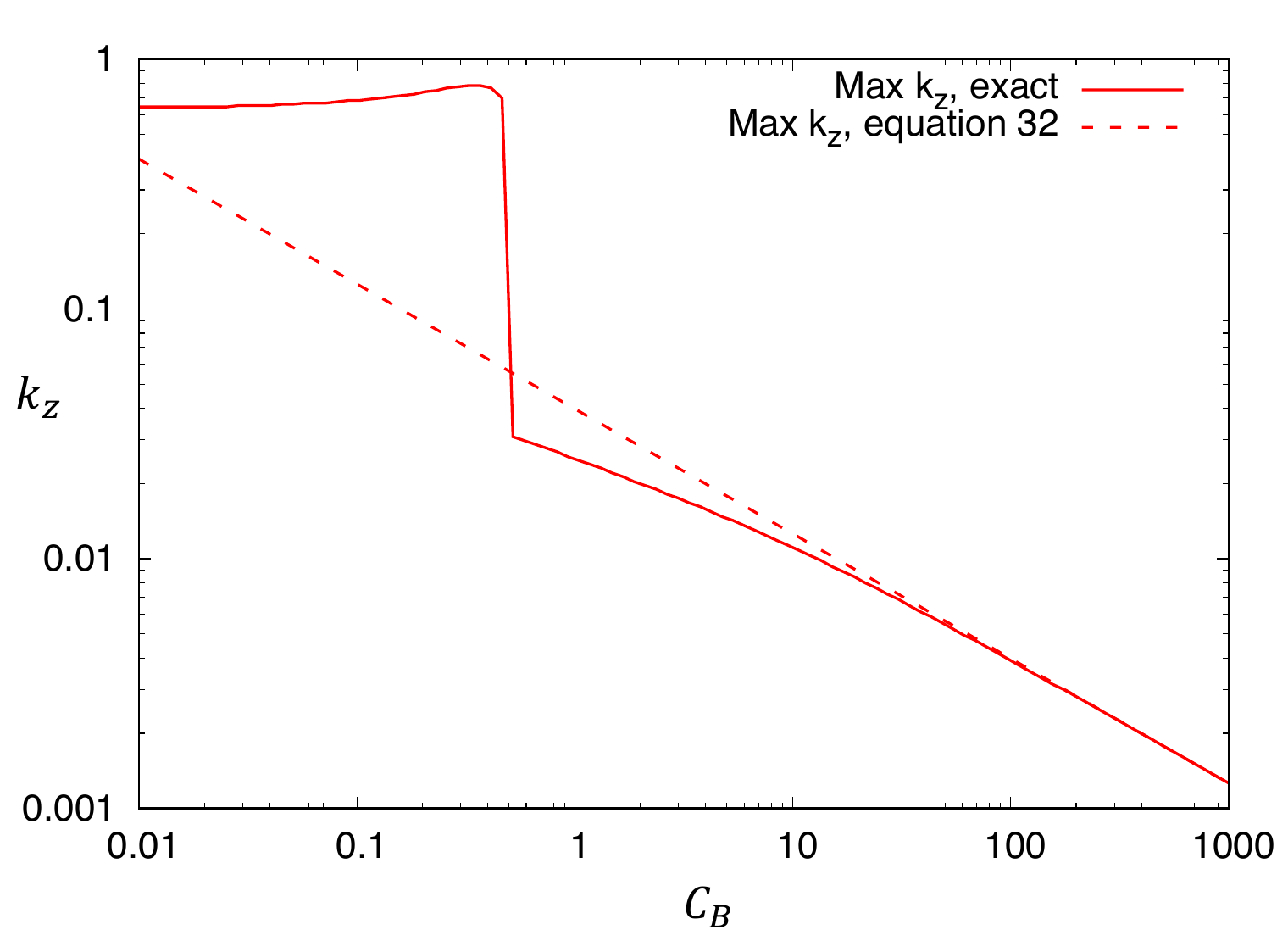}
      \caption{Fastest-growing mode wavenumber (in the reduced model) at $Re = 100$, $Rm=10$. The solid line shows the numerically-determined maximum, while the dashed line shows the solution from equation (\ref{eq:kz_approx}). The latter is a good approximation to the former for $C_B > Re$, as discussed in the main text.}
      \label{fig:asymp3}
  \end{figure}

 Finally, we can also use (\ref{eq:asymp_simple}) to find the fastest-growing mode for fixed input parameters $Re$, $Rm$ and $C_B$, by maximizing $\Re(\lambda)$ with respect to $k_z$. We obtain 
 \begin{equation}
    \kmax \simeq \pm \frac{1}{\sqrt{C_B}(Rm+Re)} \left( \sqrt{\frac{ReRm}{2}  \frac{Re-Rm }{ Re+Rm  }} -1 \right)^{1/2}.
    \label{eq:kz_approx} 
\end{equation}
after assuming that $K^2 \simeq 1$.
We see that for moderate $Re$, $\kmax = O(C_B^{-1/2} Re^{-1/2})$. Since (\ref{eq:asymp_simple}) is valid for any $k_z \ll 1/Re$, (\ref{eq:kz_approx}) is then expected to be valid whenever $C_B \gg Re$. When $C_B$ is of order $Re$ or less, (\ref{eq:kz_approx}) is not valid, and we must instead rely on numerical tools to find $\kmax$.  Figure \ref{fig:asymp3} compares the wavenumber of the fastest-growing mode obtained from 
(\ref{eq:kz_approx}) with $Re = 100$ and $Rm = 10$, with the one found by maximizing the growth rate obtained numerically over all possible values of $k_z$. We see that the approximation is quite good as long as $C_B > Re$, as expected from the analysis above. 

\section{Interpretation of the results}
\label{sec:physical_interpretation}
 
In the previous section, we demonstrated that a dramatic truncation of the full system of equations (\ref{eq:linearsys}) still retains enough physics to capture the essence of the instability studied, and that the growth rates computed from the reduced model are an excellent approximation to the true growth rate of the resistively-unstable Alfv\'{e}n waves. Moreover, we were able in some limits to obtain an analytical expression for the mode growth rate as a function of wavenumber (equation \ref{eq:asymp_simple}), which we now argue helps elucidate the mechanism responsible for the instability.

Indeed, when written non-dimensionally, equation (\ref{eq:asymp_simple}) obfuscates the physics that drive the instability. Dimensionally, however, and using the fact that $\psi_E=1/2$, 
\begin{eqnarray}
\Re(\lambda)^* \simeq  -  \frac{  (k_z^*)^2}{2  } \left( \nu^* + \eta^* \right)
+   \frac{(k_z^*)^2 (\Psi^*)^2}{ 4\nu^*} \frac{(1-Pm)}{ 1 +  \frac{(B^*)^2 (k_z^*)^2}{\rho_0^* \mu_0^* (K^*)^4} (\frac{1}{\nu^*} + \frac{1}{ \eta^*} )^2  }    
\label{eq:dimlambda_simple}
 \end{eqnarray}
 where $\Psi^* = U^*/k_x^*$ is the amplitude of the streamfunction associated with the background sinusoidal flow, and the star here denotes dimensional quantities. When $B^*$ is not {\it too} large, or when $k^*_z$ is small (i.e. when the denominator in the second term of equation \ref{eq:dimlambda_simple} is approximately equal to one), then $\Re(\lambda)^* \simeq - D^*_{\rm eff} k_z^2$, with an effective diffusivity given by:  
\begin{eqnarray}
D^*_{\rm eff} =   \frac{ 1}{2  } \left( \nu^* + \eta^* \right)
-   \frac{ (\Psi^*)^2}{ 4\nu^*} (1-Pm).     
\label{eq:Deff}
 \end{eqnarray}
 This expression is strongly reminiscent of the results obtained by \citet{DubrulleFrisch1991} who demonstrated that a {\it purely} hydrodynamic sinusoidal flow has an effective viscosity of amplitude 
\begin{equation}
    \nu^*_{\rm eff} = \nu^* - \frac{(\Psi^*)^2}{2\nu^*},  
    \label{eq:Dubrulleeq}
\end{equation}
when it acts on a cross-stream flow, which is exactly the situation we have here, albeit in the magnetized case. Note how $\nu^*_{\rm eff}$ is negative when $\Psi^* > \sqrt{2} \nu^*$, or equivalently, when $Re > \sqrt{2}$. In other words, a purely sinusoidal flows has anti-diffusive properties that can serve to {\it amplify} a cross-stream flow. 
The obvious similarities between (\ref{eq:Deff}) and (\ref{eq:Dubrulleeq}) therefore show that the resistive instability is simply caused by the well-known anti-diffusive properties of the background sinusoidal flow, acting on the slowly decaying Alfv\'en mode.

 
It is interesting to note that the $B^* \rightarrow 0$ limit of (\ref{eq:Deff}) does not recover the hydrodynamic limit of \citet{DubrulleFrisch1991}. This discrepancy has a simple physical explanation. First, note that the first term in $D^*_{\rm eff}$ is the algebraic mean of $\nu^*$ and $\eta^*$, rather than $\nu^*$ alone, because the cross-flow that is being amplified is an Alfv\'en wave at the lowest order in $\epsilon$, whose energy is therefore equally partitioned into a kinetic component, that dissipates viscously, and a magnetic component, that dissipates resistively. Second, we also see that the second term in $D^*_{\rm eff}$ does not recover $(\Psi^*)^2/2\nu^*$ when $B^* \rightarrow 0$, even when $Pm \rightarrow 0$ (i.e. when $\eta \gg \nu$, where the field and the flow merely slip with respect to one another). To see why this is the case, note that the term $(\Psi^*)^2/2\nu^*$ comes from the Reynolds stresses associated with the perturbations in the purely hydrodynamic limit, while in the MHD case, Maxwell stresses also participate in the momentum balance and partially (but not completely) cancel out the Reynolds stress. 

\section{Numerical simulations}
\label{sec:simulations}

While linear stability analyses can predict 
the initial response of a fluid in 
equilibrium to infinitesimal perturbations, they provide no immediate insight into its  
nonlinear evolution. 
In this section, we present results from a direct numerical simulation of this system for one set of physical parameters. 
It is not our intention to perform a comprehensive scan of parameter space, but instead, to provide some validation of the linear stability analysis of Sec.~\ref{sec:linearresults}, and to find out, at least qualitatively, 
how the resistively-unstable Alfv\'{e}n waves 
saturate. For this reason, 
we use the physical parameters $C_B = 1$, $\Rey = 100$, and $\Pm = 0.1$, ensuring 
that they are the only unstable modes present in the system (see Figure \ref{fig:growthrate_3Dscan}).

We use the pseudospectral code Dedalus \citep{Dedalus} to 
evolve a two-dimensional version of Eqs.~\eqref{eq:normalized_full_PDEs} (expressed in terms of a streamfunction $\psi$ and flux function $A$) in time. The simulation is initialized with a unit-amplitude sinusoidal shear flow $\mathbf{u} = \mathbf{u}_E$ and a uniform magnetic field $\mathbf{b} = {\bf b}_E = {\bf e}_z$, i.e., Eqs.~\eqref{eq:u_E} and \eqref{eq:b_E}. We note that this is an equilibrium solution, as the forcing term $-\Rey^{-1}\nabla^2\mathbf{u}_E$ in Eq.~\eqref{eq:normalized_full_PDEs} balances the viscous dissipation of $\mathbf{u}_E$. We perturb this equilibrium by adding 
small-amplitude white noise 
to the streamfunction, which
seeds the instability. 
The simulation uses a domain size of $(L_x, L_z) = (4\pi, 170\pi)$, where the dimensions are chosen to accommodate two wavelengths of the sinusoidal equilibrium in the $x$ direction and approximately two wavelengths of the fastest-growing mode, according to the linear stability analysis, in the $z$ direction. The domain is doubly-periodic, with a resolution of
64 Fourier modes in the $x$ direction and 256 modes in the $z$ direction. Nonlinear terms are dealiased using the standard 3/2-rule, and we use a four-stage, third-order, implicit-explicit Runge-Kutta timestepping scheme to evolve the solution in time \citep[][Sec.~2.8]{ascher}.

\begin{figure}
    \centering
    \includegraphics[width=\textwidth]{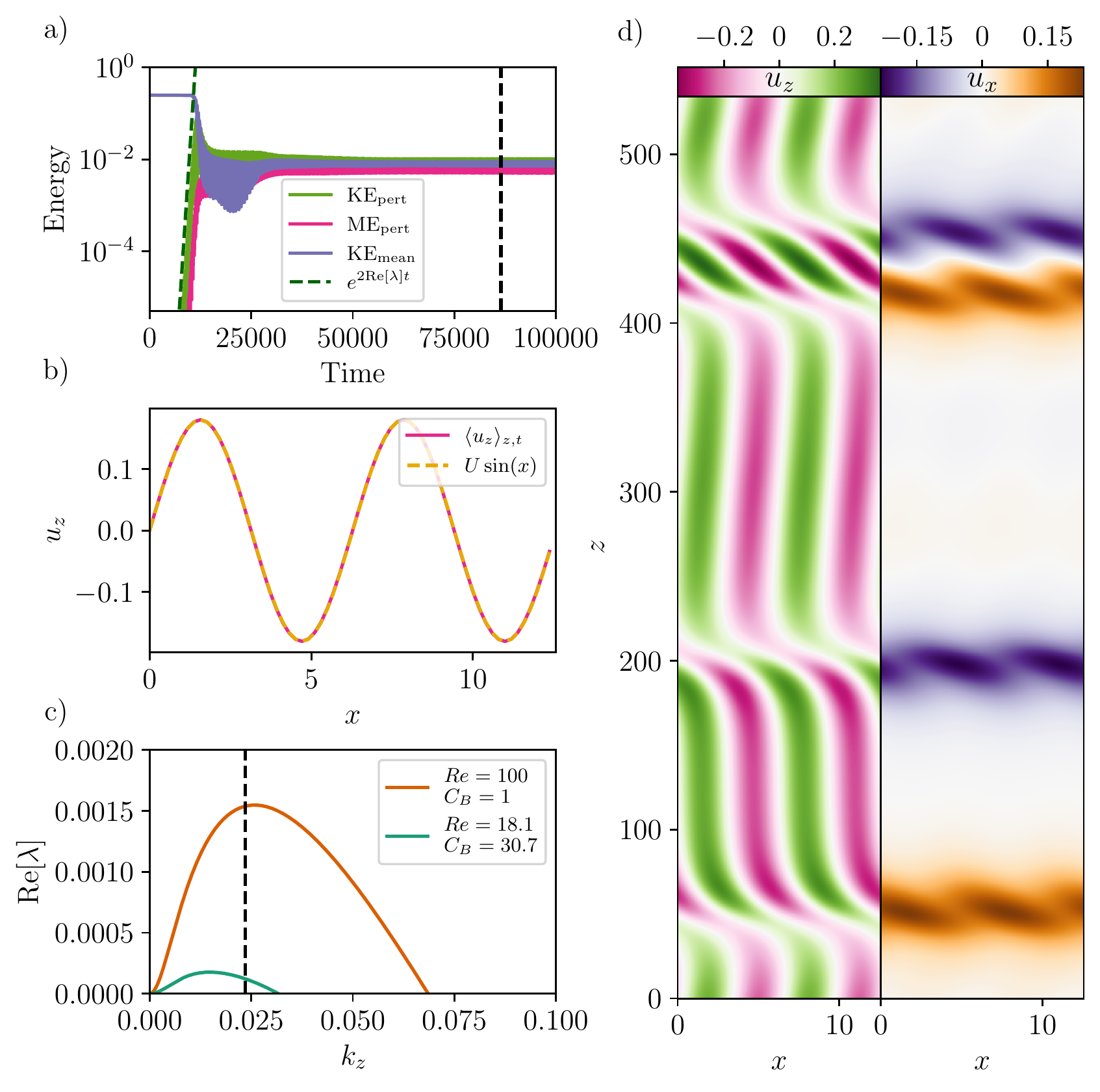}
    \caption{Panel a): kinetic and magnetic energies of perturbations about the mean are shown alongside the kinetic energy of the mean flow for a simulation with $C_B = 1$, $\Rey = 100$, and $\Pm = 0.1$, with the green dashed line showing the growth rate of the most-unstable mode that fits into this domain size according to Sec.~\ref{sec:linearresults}, demonstrating consistency with the results of that section. Panel b): the vertical mean flow, averaged in both $z$ and $t$ (where the time-average is taken over the second half of the simulation), is shown with a sine wave overplotted to demonstrate how nearly sinusoidal the flow remains. Panel c): the dispersion relation based on the initial values of $\Rey$ and $C_B$ (orange) is shown alongside the dispersion relation based on the values of $\Rey_{\rm eff}$ and $C_{\rm eff}$ achieved in saturation (green, see text), with the black vertical line showing the wavenumber corresponding to a wavelength that equals the domain height, demonstrating that even in saturation, the mean flow profile remains linearly unstable. Panel d): a snapshot of the vertical (left) and horizontal (right) flow is shown at $t \approx 87000$ (indicated by the vertical dashed line in panel a). At this time, the system is dominated by two counter-propagating solitons.}
    \label{fig:DNS}
\end{figure}

The results are summarized in Fig.~\ref{fig:DNS}. We decompose the flow according to $\mathbf{u} = \langle \mathbf{u} \rangle_z + \mathbf{u}'$, where $\langle \cdot \rangle_z$ denotes an average in $z$, and refer to $\langle \mathbf{u} \rangle_z$ as the mean flow and $\mathbf{u}'$ as the fluctuating flow, with analogous definitions for the mean 
and fluctuating magnetic field. Note that the mean 
flow remains purely vertical, i.e., $\langle \mathbf{u} \rangle_z = \langle u_z \rangle_z \mathbf{e}_z$. The kinetic and magnetic energies of the fluctuations are shown alongside the kinetic energy of the mean 
flow in Fig.~\ref{fig:DNS}a. As expected, the fluctuations grow early in the simulation at a rate that is consistent with the linear stability results of Sec.~\ref{sec:linearresults}, shown by the dashed green line. 
Also consistent with Sec.~\ref{sec:linearresults} is the dominance of kinetic energy over magnetic energy of fluctuations in the linear growth phase, see Fig.~\ref{fig:kmax_energy_scan}. 
Finally, snapshots of the flow and field perturbations at this stage (not shown) 
are also consistent with those of the unstable modes shown in Fig.~\ref{fig:resistive_structures}.

The exponential 
growth phase of the resistive Alfv\'en modes ends around $t = 10000$, when their amplitudes become commensurate with that of the mean 
flow. 
Nonlinear interactions then cause a rapid decrease in the amplitude of the mean 
sinusoidal flow, from an original value of one down to an average of about $U = 0.181$ (see Fig.~\ref{fig:DNS}b). The shape remains almost exactly sinusoidal, however. One may therefore wonder whether, by reducing the mean 
flow amplitude, the system has simply adjusted itself in such a way as to become marginally stable to all modes of instability, which is a common route towards saturation. To test this idea, we use the new flow amplitude $U$ to compute effective Reynolds numbers  $Re_\mathrm{eff} = U Re \simeq 18.1$ and $Rm_\mathrm{eff} = U Rm \simeq 1.81$, and an effective magnetic parameter $C_\mathrm{eff} = C_B / U^2 = 30.7$. 
Using these effective parameters, we can compute the growth rate of unstable modes on the new background flow, and the results are shown in Fig.~\ref{fig:DNS}c (green curve), together with a dashed vertical line indicating the wavenumber of a mode whose wavelength equals the domain height. We clearly see that the system remains unstable to domain-size resistive Alv\'en modes, showing that the path to saturation suggested above is not relevant for this simulation. 

Furthermore, inspection of the shape and relative amplitude of the flow and field perturbations in the nonlinear state reveals that they are profoundly different from those of linear modes computed in Section \ref{sec:linearresults}. Indeed, 
Fig.~\ref{fig:DNS}d shows $u_z$ and $u_x$, the vertical and horizontal components of the flow (including mean and fluctuations), at $t \approx 86480$, as indicated by the dashed vertical line in the top-left panel. 
While the unstable modes calculated in Sec.~\ref{sec:linearresults} are sinusoidal in $z$, the saturated state is dominated by fluctuations that are localized to narrow 
 structures in the $z$ direction. These structures appear to be two pairs of counter-propagating solitons, most easily seen in terms of the localized horizontal flows shown in the $u_x$ snapshot. The positive-$u_x$ fluctuations propagate in the $+z$ direction, while the negative-$u_x$ fluctuations propagate in the $-z$ direction at the same speed, with each soliton unperturbed as it travels through a counter-propagating soliton.
 
 Finally, note that we have run many other nonlinear simulations of the  resistively-unstable Alfv\'en modes for a variety of input parameters (not shown here). Similar solitons appeared in all cases. The simplicity of these dynamics suggests they may be well-described by a weakly nonlinear analytical model. Such a model, as well as an exploration of how these solitons vary for different physical parameters, domain sizes, and three-dimensional systems, is left to future work.  

\section{Discussion and conclusion}
\label{sec:conclusions}
We have investigated the linear stability of a sinusoidal shear flow with an initially uniform, streamwise magnetic field in two-dimensional, incompressible MHD with finite viscosity and resistivity. We found three modes of instability, unlike the single KH mode present for this flow in ideal MHD or in the absence of a magnetic field. One of these modes corresponds to the usual KH mode, while the other two modes, to our knowledge, have not been identified elsewhere in the literature. This paper focused on understanding the dynamics of one of these new modes, which we refer to as resistively-unstable Alfv\'{e}n waves. These modes appear as pairs of counter-propagating unstable waves and exist for all magnetic field strengths, but only when the magnetic Prandtl number $\Pm < 1$. By deriving a reduced model for this particular mode of instability, we were able to show that it is amplified by the negative eddy viscosity of periodic shear flows identified by \cite{DubrulleFrisch1991}. Finally, we presented a direct numerical simulation of the nonlinear evolution of these waves, demonstrating that they saturate in a quasi-stationary state dominated by counter-propagating solitons.

The physical parameters for which this new mode of instability exists leads to two significant consequences worth stressing. First, while the resistively-unstable Alfv\'en waves require finite dissipation (with $\Rey$ and $\Rm$ above a certain threshold), we found that they remain unstable no matter how large $\Rey$ and $\Rm$ become, provided $\Pm < 1$. As a consequence, even when modeling astrophysical plasmas with extreme Reynolds numbers, calculations that employ ideal MHD may erroneously neglect this instability. Second, unlike the ordinary KH mode found in ideal MHD (or its counterpart in this system) which becomes stable for sufficiently strong magnetic fields, the resistively-unstable Alfv\'en waves are unstable for all nonzero magnetic field strengths. Thus, counter to common intuition that shear-flow instabilities are stabilized by parallel magnetic fields of sufficient strength \citep{chandrasekhar}, our results demonstrate that, at least for the flow profile considered here, instability can persist for arbitrarily large magnetic field strengths. 

As shown in Sec.~\ref{sec:physical_interpretation}, the underlying mechanism for this instability stems from the negative eddy viscosity of periodic shear flows described by \citet{DubrulleFrisch1991}, which amplifies the shear Alfv\'{e}n waves present in this system in the absence of the background shear flow. The simplicity of this mechanism suggests it might be quite general, and exist in other scenarios where a sinusoidal shear flow is added to a system that naturally supports waves transverse to the mean flow. Indeed, \citet{garaud_stratified_kolmogorov} similarly found low-wavenumber oscillatory modes when studying sinusoidal shear flow in a stratified fluid, that only exist when viscosity is taken into account. While a detailed investigation of these modes was beyond the scope of that work, they appear similar to the ones reported here, with internal gravity waves in that system playing the role of Alfv\'{e}n waves in the MHD system. We speculate that similar modes might exist in reduced plasma models that permit zonal flows and drift waves \citep[e.g.,][]{zhu}.

We envision two primary directions for future work based on these results. The first is a thorough investigation of the nonlinear evolution of this instability. We have demonstrated for one set of parameters that this system saturates in a quasi-stationary state that supports counter-propagating solitons. Additional simulations performed over a broad range of physical parameters will be needed to characterize how the speed, number, and shape of these solitons vary with input parameters. Furthermore, the simple nature of this saturated state invites efforts to develop reduced nonlinear models that can be compared against simulations. Finally, even though we demonstrated that 2D modes of instability are the fastest-growing ones in the linear regime, it is likely that the saturation of the instability will be profoundly different in 2D and 3D, and it is unclear whether these solitons will persist.

The second direction for future work is to explore the physical implications of these resistively-unstable Alfv\'{e}n waves. As described in Sec.~\ref{sec:intro}, 
the double-diffusive fingering instability drives ``elevator" modes that flow in the vertical direction and vary sinusoidally in the horizontal directions. Their saturation is traditionally modeled by requiring a balance between the finger growth rate, and the growth rate of parasitic shear instabilities \citep{RadkoSmith,Brown}. \citet{harrington_enhanced_2019} (hereafter HG19) recently studied the effect of an added vertical magnetic field, demonstrating both numerically and theoretically that the latter decreases the shear instability growth rate and therefore strongly affects the saturation of the fingering instability. However, 
all of their simulations were performed with $\Pm = 1$, and their shear-flow stability analysis assumed ideal MHD; thus, the effects of resistively-unstable Alfv\'{e}n waves were not present in their simulations or in their reduced model. Since the stellar interiors where fingering convection occurs generally satisfy $\Pm < 1$, 
it is possible that the newly discovered modes have an effect on the saturation of magnetized fingering convection that is not accounted for by HG19. 


Finally, note that all of the results presented in this paper were obtained for a sinusoidal flow that varies in only one of the two horizontal directions -- a planar shear flow -- which is the geometry for which the \citet{DubrulleFrisch1991} mechanism was originally discussed. However, in many of the instabilities discussed above, the primary elevator modes vary sinusoidally along both horizontal axes, as seen in Fig.~1 of \citet{harrington_enhanced_2019} or discussed in Sec.~3.2 of \citet{RadkoSmith}. It will therefore be important to establish in future work whether the instability mechanism discovered here remains active for MHD shear flows where the shear varies along two axes, e.g. for flows with structure $\mathbf{u}_E = \sin(x)\sin(y) \mathbf{e}_z$.

\backsection[Acknowledgements]{We acknowledge use of the Lux supercomputer at UC Santa Cruz, funded by NSF MRI grant AST-1828315. This work used the Extreme Science and Engineering Discovery Environment (XSEDE) Expanse supercomputer at the San Diego Supercomputer Center through allocation TG-PHY210050. 
We thank Steve Tobias, David Hughes, Paul Terry, Ellen Zweibel, MJ Pueschel, Noah Hurst, Alexis Kaminski, and Jeff Oishi for useful conversations.}


\backsection[Funding]{AF and PG acknowledge support from the National Science Foundation grant AST-1908338. IC acknowledges the support of the University of Colorado’s George Ellery Hale Graduate Student Fellowship. Some of this work was part of a Kavli Summer Program in Astrophysics project, funded by the Kavli Foundation.}

\backsection[Declaration of interests]{The authors report no conflict of interest.}

\backsection[Data availability statement]{The simulation data presented here is available upon reasonable request.}

\backsection[Author ORCID]{A.E.~Fraser, https://orcid.org/0000-0003-4323-2082; I.G.~Cresswell, https://orcid.org/0000-0002-4538-7320; P.~Garaud, https://orcid.org/0000-0002-6266-8941.}

\bibliographystyle{jfm}
\bibliography{resistive}

\end{document}